\documentclass[aps,amssymb,superscriptaddress,amsmath,twocolumn,showkeys,nofootinbib,pra]{revtex4-1}

\usepackage{graphicx}
\usepackage{epstopdf}
\usepackage{bm}
\usepackage[all]{xy}
\usepackage{afterpage}
\usepackage{float}

\usepackage{microtype}

\begin{document}

\title{Dynamical Consequences of Time-Reversal Symmetry for Systems with Odd Number of Electrons: Conical Intersections, Semiclassical Dynamics, and Topology
}

\author{Ruixi \surname{Wang}}
\affiliation{Department of Chemistry, Wayne
State University, 5101 Cass Ave, Detroit, MI 48202}
\author{Vladimir Y. \surname{Chernyak}}
\affiliation{Department of Chemistry, Wayne
State University, 5101 Cass Ave, Detroit, MI 48202}
\affiliation{Department of Mathematics, Wayne
State University, 656 W. Kirby, Detroit, MI 48202}
\date{\today}

\begin{abstract}
In this manuscript we identify the main differences between the effects of Kramers symmetry on the systems with even and odd number of electrons, the ways how the aforementioned symmetry affects the structure of the Conical Seams (CSs), and how it shows up in semiclassical propagation of nuclear wavepackets, crossing the CSs. We identify the topological invariants, associated with CSs, in three cases: even and odd number of electrons with time-reversal symmetry, as well as absence of the latter. We obtain asymptotically exact semiclassical analytical solutions for wavepackets scattered on a CS for all three cases, identify topological features in a non-trivial shape of the scattered wavepacket, and connect them to the topological invariants, associated with CSs. We argue that, due to robustness of topology, the non-trivial wavepacket structure is a topologically protected evidence of a wavepacket having passed through a CS, rather than a feature of a semiclassical approximation.
\end{abstract}

\maketitle

\section{Introduction}
\label{sec:intro}

Nonadiabatic effects play a crucial role in photoinduced molecular dynamics in condensed, as well as gas phase, in small and large molecules, showing up in various kinds of photoinduced reactions \cite{nitzan2006chemical,doi:10.1002/anie.198302101,doi:10.1021/cr2001299}, including photo-dissociation \cite{doi:10.1063/1.2203611,Kim1561} and photo-isomerization \cite{SEIDNER199427,doi:10.1146/annurev.physchem.57.032905.104612}. Conical Intersections (CIs) \cite{domcke2004conical,Matsika2011,Domcke2012} play an extremely important role in all areas that involve non-adiabatic transitions by a variety of reasons. First of all they are unavoidable in a sense that once you have a single CI point, by the implicit function theorem \cite{krantz2012implicit}, you immediately get a codimension $2$, i.e., $(d-2)$-dimensional Conical Seam (CS). In fact it is natural to interpret an avoided crossing, i.e., two adiabatic Potential Energy Surface (PES) coming close, as the system just being close to an unidentified (e.g., due to considering a reduced configuration space) CS. Secondly, CIs have been found computationally in a variety of molecules \cite{Yarkony1996a,doi:10.1063/1.3028049,doi:10.1021/jp0633897}, by using clever identifying algorithms \cite{doi:10.1021/ct7002435,RAGAZOS1992217}, as well as multi-reference electronic structure methods \cite{doi:10.1021/jp0633897,Matsika2003a,doi:10.1021/jp0761618} capable of handling the symmetries, associated with electronic state degeneracy. Thirdly, CI are usually kept responsible for ultrafast nonadiabatic transitions in gas phase \cite{doi:10.1021/jp054158n,doi:10.1021/ja0258273,doi:10.1021/ja904780b}, as well as ultrafast photo-relaxation \cite{doi:10.1021/jp048284n,C6CP01391K} and photo-isomerization \cite{KIM201286,JIANG2012107} in biological molecules, including such super-important examples as photo-isomerization of rhodopsin \cite{HAHN2000297,Schoenlein412,doi:10.1063/1.2034488}.

CIs have been under theoretical/computational studies in terms of their electronic structure \cite{doi:10.1063/1.3028049,doi:10.1063/1.462715,domcke2011conical}, as well as wavepacket dynamics in both closed \cite{doi:10.1063/1.2049250,NIKOOBAKHT2016221,Galbraith2017} and open (a molecule coupled to environment/heat bath) \cite{doi:10.1063/1.467618,doi:10.1063/1.1423326,C6FD00088F} cases. The dynamical problem, however, is complex for a full quantum treatment, even assuming one has perfect knowledge on PES and non-adiabatic couplings, due to high dimensionality of the configuration space, even in the closed dynamics case. Even sophisticated computational schemes, capable of performing quantum mechanical simulations \cite{THOSS2006210,C5CP07332D,BECK20001}, including such clever approaches as spawning \cite{doi:10.1021/ar040202q,doi:10.1021/jp994174i}, scale unfavorably with the number of atoms, so that much more cost-efficient semiclassical or/and mixed quantum-classical methods are highly desired.

There is, however, an intrinsic issue on the way of developing semiclassical approaches capable of handling non-adiabatic transitions in a proper way. As long as adiabatic dynamics is concerned, there is an excellent understanding of the semiclassical limit, which is asymptotically exact, in terms of the semiclassical Van-Vleck \cite{schulman2012techniques,gutzwiller2013chaos}, or more sophisticated Herman-Kluk \cite{HERMAN198427,doi:10.1063/1.450142}, providing an intuitive picture for the case when semiclassical propagation is no longer quantitatively precise. Such intuition is still to merge for the case of non-adiabatic transitions, despite a several decades activity in the aforementioned field. There is a reason behind that. Although apparently not truly appreciated in modern literature on non-adiabatic transitions, it is known from 1950s \cite{landau2013quantum} that in the absence of level crossing, in the truly semiclassical $\hbar \to 0$ limit, the nonadiabatic effects vanish in a non-analytical exponential $\sim \exp (- 1/\hbar)$ way; therefore applying semiclassical approximations that involve hopping far from true intersections, in a situation when the semiclassical parameter is not small faces serious difficulties.

The surface hopping algorithms originally formulated by Tully \cite{TullyJohn,Tully1990} and further developed by other authors \cite{Mario,doi:10.1063/1.3506779}, that scale favorably with the system size, provide an efficient tool for studying non-adiabatic effects in molecular dynamics \cite{nitzan2006chemical,Mario,doi:10.1021/jp109522g}, especially in condensed phase \cite{C2CP40300E,NIEBER2008405}. The aforementioned algorithms generally address the problem of branching ratios and demonstrate an ability to predict/interpret the latter with decent accuracy. However, due to extremely intuitive nature of these algorithms (namely the associated Monte Carlo procedure does not converge to the solutions of the Schr\"{o}dinger equation), as well as not accounting for the wavefunction phase, associated with the classical trajectories in between the hopping events, makes their capability to predict the wavepacket shapes a big question. More sophisticated algorithms, see e.g.,~\cite{Gorshkov2013,White2014}, that represent an attempt to solve the Schr\"{o}dinger equation using a valid Monte Carlo scheme, faces a problem of making a choice which integrations in a path-integral representation should be done in the saddle-point, and which should be performed numerically exactly using a Monte Carlo procedure. The problem is usually referred to as the choice of initial conditions for the classical trajectory after hopping \cite{PhysRev.181.166,PhysRev.181.174}, with the numerical results being choice dependent, which demonstrates inconsistency of the scheme itself. Also the method apparently has not shown capability of describing non-adiabatic dynamics near CSs.

The wavepacket shapes are very important for interpreting certain experimental data, e.g., photo-dissociation in gas phase \cite{doi:10.1063/1.481130,doi:10.1021/jacs.6b03288}. In particular, Picconi and Grebenschikov recently clearly demonstrated, using state-of-the-art numerically exact solution of the dynamical Schr\"{o}dinger equation, that certain characteristic features, acquired by wavepackets after passing through the conical seams \cite{C5CP04564A,doi:10.1063/1.5019735,doi:10.1063/1.5019738}, clearly show up in the experimental photo-dissociation data. Mukamel with co-authors \cite{kowalewski2017monitoring,doi:10.1021/acs.chemrev.7b00081} recently proposed novel time-resolved X-ray experiments, with the data being sensitive not only to the shapes of the scattered wavepackets, but also to its evolution when passing the CSs, thus providing detailed information on the local structure of the latter.

Fortunately, there is a well-defined, i.e., an asymptotically exact in the $\hbar \to 0$ limit, semiclassical approximation scheme that accounts for non-adiabatic transitions. It has been formulated, including introducing a dimensionless parameter $g_{\rm s}$ that characterizes the validity of the approximation, in~\cite{Piryatinski2005} on a level of modifying the Van Vleck semiclassical propagator, using the solution of the celebrated Landau-Zener problem in~\cite{landau1932theorie,zener1932non} when a classical trajectory passes nearby a CS. Several authors have been using similar approaches on an intuitive level \cite{C5CP04564A,MALHADO200839} before and after the scheme was rigorously formulated \cite{Piryatinski2005}. The aforementioned method/methods is free from the $\sim \exp (- 1/\hbar)$ scaling problem, mentioned above, since it simply states that there are no non-adiabatic transitions, unless the trajectory is close to a CS. Notably, it has been shown in~\cite{C5CP04564A} that the aforementioned semiclassical approach shows excellent agreement with numerically exact results for realistic systems. This is the approach we adopt in this paper to extend the method to the half-integer spin (odd number of electrons) situation.

Unavoidable crossing for systems with odd number of electrons is much less studied. To our best knowledge the problem was addressed by Mead \cite{Mead1979,Mead1987}, who pioneered the study of the local structure of CS in the aforementioned case by accounting for time-reversal symmetry that leads to Kramers permanent $2$-fold degeneracy of electronic levels, and established the codimension of the CSs to be $5$, as opposed to its value $2$ for the ''standard'' case of integer-spin systems, followed by~\cite{Pacher1989}, where an ${\rm SU} (2)$ non-abelian gauge field has been introduced, that can be, with minimal abuse of notation, referred to as the diagonal (i.e., acting within the same double-degenerate PES) components of the non-adiabatic coupling terms. The ideas of Mead were further developed by Matsika and Yarkony in a series of papers \cite{doi:10.1063/1.1378324,doi:10.1063/1.1391444,doi:10.1063/1.1427914,doi:10.1021/jp020396w} in the early 2000s, including developing algorithms capable of identifying the CSs. In the integer spin case, CSs bring in a well-known topological effect, which is often referred to as geometrical Berry phase, which is in fact topological, since the phase admits discrete values $0$ and $\pi$, meaning that an adiabatic state acquires a sign factor of $-1$ upon going around the CS. The topological Berry phase, associated with a CS has dynamical implications, namely the parts of a wavepacket component, staying on the same adiabatic surface, that pass the CS on its different sides, acquire an additional sign factor $\pm 1$, modifying the wavepacket shape. The above effect has been discussed in, e.g.,~\cite{Piryatinski2005}, in the semiclassical approximation.

The goal of the presented study is to identify the implications of time-reversal symmetry and topology, associated with unavoidable crossings, on wavepacket dynamics, with focus on half-integer spin case. This is achieved via (i) investigating electronic structure using group invariance under real structure transformations, (ii) establishing the topological invariants, associated with a CS, (iii) extending the semiclassical approach of~\cite{Piryatinski2005} to the half-integer spin case, characterized by the Kramers permanent degeneracy, by addressing the Kramers permanent degeneracy for the adiabatic Born-Oppenheimer dynamics, as well generalizing the ballistic approach that describes the wavepacket evolution when passing a CS, and (iv) finally identifying the topological implications on the scattered wavepacket shape.


Topology is involved via establishing a topological obstruction for building a global adiabatic basis set on a $4$-dimensional sphere, surrounding a CS with codimension $5$, and identify it with the second Chern class \cite{milnor2016characteristic} that has integer nature, as opposed to its integer counterpart, characterized by a binary, i.e., $\pm 1$ factor-like Stiefel-Whitney class. Since a CS is surrounded by a $4$-dimensional sphere rather than a circle, as in the integer-spin case, non-trivial topology around a CS is not related to any topological Berry phase.
However, on the path to extending the Born-Oppenheimer scheme to the permanent degeneracy case, we demonstrate that effects, related to geometrical (i.e., admitting continuous values) Berry phase, namely parallel transport, show up in semiclassical evolution of the wavefunction polarization, defined as a unit complex $2$-component vector that describes the vector character of the wavefunction, associated with permanent degeneracy, and therefore resides in a $3$-dimensional sphere. Due to the aforementioned vector character, the geometrical Berry phase becomes non-abelian, i.e., is represented by an element of the special unitary group ${\rm SU}(2)$ that describes the rotation, acquired by the wavefunction polarization upon a nuclear configuration going along a closed path (loop). In particular the Berry phase effect is associated with diagonal (i.e., acting within the same $2$-fold degenerate adiabatic surface) components of the non-adiabatic coupling terms~\cite{Pacher1989}, represented by a non-abelian ${\rm SU} (2)$ gauge field \cite{faddeev2018gauge} with non-zero curvature/intensity.


The manuscript is organized as follows. In section~\ref{sec:kramers-symmetry} we discuss the time-reversal symmetry for non-relativistic (with spin-orbit corrections) many electron systems, with focus on the half-integer spin (odd electron number) case, and the major differences between the aforementioned situation and the integer spin or no time-reversal symmetry (strong effects of magnetic fields) counterparts. To provide a formulation, ready for studies of the dynamical consequences of time-reversal symmetry, we identify the orthogonal ${\rm SO}(n)$, unitary ${\rm U}(n)$, and symplectic groups ${\rm Sp}(n)$ as the ones, responsible for the symmetry in the integer-spin, no-symmetry, and half-integer spin situations, naturally referring to them as the orthogonal, unitary, and symplectic cases. We also provide a description of local structures of CSs in all three cases, within a unique fashion, which is achieved by introducing the gamma-matrices, associated with spinors in lower an higher dimensions, namely $d=2$, $d=3$, and $d=5$. In section~\ref{sec:conical-topol-inv} we introduce the topological invariants associated with CSs for all three aforementioned cases, represented by the first Stiefel-Whitney, first Chern, and second Chern classes, respectively, focusing on their properties that unveil the topological effects in wavepacket dynamics. The presentation is done on an intuitive level, keeping it to minimum, needed to understand the topological dynamical implications, using calculus and linear algebra only, so it does not require from a reader any knowledge in topology or geometry. We also describe a geometrical effect, associated with Kramers permanent degeneracy, and relate it to ''non-abelian Berry phase''. In section~\ref{sec:BO-semicl-half-int} we extend the Born-Oppenheimer, and corresponding semiclassical approximations to the symplectic case, introduce the wavefunction polarization, appearing as a consequences of Kramers degeneracy, and apply the parallel transport concept to describe semiclassical evolution of the polarization. In section~\ref{sec:conical-semicl-half-int} we present the ballistic approximation, the main tool of studying the asymptotically exact semiclassical limit of wavepacket evolution while it passes a CS, in a unique fashion, in particular allowing to apply the standard expressions of the celebrated $2$-state Landau-Zener problem to its $5$-state counterpart, occurring in the symplectic case, which is achieved by making use of the (Clifford) algebra of gamma-matrices, which, in particular, justifies the formulation of time-reversal symmetry, given in section~\ref{sec:kramers-symmetry}. In section~\ref{sec:toplology-scatrd} we present an explicit analytical expression for the wavepacket that just passed through a conical seam, analyze its shape, identify the topological nature of the latter, and connect it to the topological invariants of CSs, described in section~\ref{sec:conical-topol-inv}.

Not to distract a reader form the main presentation flow, certain details are presented in the appendices. In appendix~\ref{sec:Sp-groups} we put some technical details, associated with the symplectic case of time-reversal symmetry. One of the reasons, we wrote this appendix is that multiple notation is used in literature on symplectic groups. In appendix~\ref{sec:spinors}, since gamma-matrices, associated with different dimensions play an important role in our approach, and also for the sake of completeness we put some basic facts about spinors and gamma-matrices in an arbitrary dimension. 
Finally, in appendix~\ref{sec:diff-forms-Chern-cl} we present, in a self-sufficient way, some basic facts on differential forms, including wedge products and multidimensional Stokes theorem, as well as representation of Chern classes using differential forms, in order to support some intuitive arguments on topological properties/invariants, associated with the CSs and presented in section~\ref{sec:conical-topol-inv} with more formal derivations.

\section{Time-Reversal Symmetry in Many-Electron Systems}
\label{sec:kramers-symmetry}

Time-reversal symmetry that occurs in the absence of external magnetic fields, which is usually the case in dynamics of molecular systems, has important implementations on the system dynamics. In the simplest case of no spin the time reversal transformation $j$ is defined by $j\psi(\bm{r})= \psi^{*}(\bm{r})$, and it commutes $j\hat{H}= \hat{H}j$ with the system Hamiltonian $\hat{H}$. Such symmetry is coined time-reversal due to the fact that if $\psi(t)$ satisfies the dynamical Schr\"{o}dinger equation $i\hbar \partial_{t}\psi(t)= \hat{H}\psi(t)$ then, due to the above commutation property $j\psi(t)= \psi^{*}(t)$ satisfies the Schr\"{o}dinger equation with the same Hamiltonian, but for reversed time.

Since, by simple intuitive reason, presented above, time-reversal symmetry involves complex conjugation, it is represented by an {\it antilinear} map $j$ acting in the space of the system quantum states, which means
\begin{eqnarray}
\label{anti-linear} j(u+ v) = j(u)+ j(v), \;\;\; j(\lambda u)= \lambda^{*}j(u)
\end{eqnarray}
for any two states $u, v$ and complex number $\lambda$. Note that antilinear is different from a ''standard'' linear map through the second condition that in the linear case map reads $j(\lambda u)= \lambda j(u)$.

Time-reversal symmetry for spin $1/2$ has been identified by Kramers, and can be described in the following way. One can ask a question: How does an antilinear map $j$, referred to as a real structure, look like that acts in a $2$-dimensional complex vector space of the spin $1/2$ and commutes with the group $SU(2)$ action? The latter property is equivalent to commuting with the generators of the corresponding Lie algebra $su(2)$, represented by $-i\bm{\sigma}= -(i\sigma_{x}, i\sigma_{y}, i\sigma_{z})$, with
\begin{eqnarray}
\label{Pauli-matrices} && \sigma_{x}= \sigma_{1} = \left(\begin{array}{cc} 0 & 1 \\ 1 & 0 \end{array}\right), \;\;\; \sigma_{y} = \sigma_{2} = \left(\begin{array}{cc} 0 & -i \\ i & 0 \end{array}\right), \nonumber \\ && \sigma_{z} = \sigma_{3} = \left(\begin{array}{cc} 1 & 0 \\ 0 & -1 \end{array}\right), \;\;\; \sigma_{0} = I = \left(\begin{array}{cc} 1 & 0 \\ 0 & 1 \end{array}\right)
\end{eqnarray}
being the Pauli matrices in on of their standard representations. A real structure $j$ that satisfies the aforementioned commutation property is represented by a matrix
\begin{eqnarray}
\label{real-structure-spin-1/2} j = \eta i\sigma_{y}= \left(\begin{array}{cc} 0 & \eta \\ -\eta & 0 \end{array}\right), && \;\;\;\; j \left(\begin{array}{c} c_{1} \\ c_{2} \end{array}\right) =  \left(\begin{array}{c} \eta c_{2}^{*} \\ - \eta c_{1}^{*} \end{array}\right) \nonumber \\ && j^{2}= -1.
\end{eqnarray}
with $\eta \in {\rm U}(1)$ being a unimodular factor. The commutation properties follow from the commutation relations $\sigma_{x}\sigma_{y}= -\sigma_{x}\sigma_{y}$, $\sigma_{y}\sigma_{z}= -\sigma_{z}\sigma_{y}$, and $\sigma_{z}\sigma_{x}= -\sigma_{x}\sigma_{z}$, combined with the anti-linearity of $j$. The real structure, defined by Eq.~(\ref{real-structure-spin-1/2}) possesses two important properties: it preserves scalar products in the sense
\begin{eqnarray}
\label{preserve-scalar} (j(u),j(v))= (u,v)^{*},
\end{eqnarray}
and $j^{2}= -1$.
A straightforward argument that involves the Schur's lemma shows that Eq.~(\ref{real-structure-spin-1/2}) completely classifies the real structures with $j^{2} = -1$ that preserve the scalar product. Hereafter we choose $\eta = 1$. It is straightforward to verify that the Breit-Pauli Hamiltonian \cite{bethe2012quantum} commutes with the real structure $j$ obtained by applying $j$, defined by Eq.~(\ref{real-structure-spin-1/2}), to the spin variables of all electrons. An obvious, bur extremely important consequence of the property of the spin $1/2$ real structure [see Eq.~(\ref{real-structure-spin-1/2})] is $j^{2}= (-1)^{N}$ with $N$ being the number of electrons, so that we have $j^{2}= 1$ and $j^{2}= -1$ for the even and odd number of electrons, respectively, which leads to very different electronic structure symmetry properties for molecules and radicals.

To describe a situation when a finite number of Potential Energy Surfaces (PES) are taken into account we should consider an $n$-dimensional complex vector space $V$ (or, equivalently, a $2n$-dimensional real vector space) that describes the space of electronic states for a given nuclear configuration, $n$ being the number of PES, taken into consideration, with an action $j:V \to V$ of a real structure on it, and further identify the space of allowed electronic Hamiltonians, represented by Hermitian operators $h$ acting in $V$ and commuting with $j$.

To recover a well-known picture, we start with the $j^{2}= 1$ case that corresponds to an even number of electrons. The analysis is very simple: we consider $V$ as a $2n$-dimensional real vector space with $i: V \to V$ representing multiplication with the imaginary unit $i$. Note that $i$ and $j$ can be viewed as just linear maps acting in the $2n$-dimensional real vector space $V$. Since $j$ preserves the scalar product in the real space it is an orthogonal operator. Generically an orthogonal operator has pairs of mutually complex conjugated eigenvalues, however, due to the $j^{2}= 1$ condition, all eigenvalues are $\pm 1$ and hence $j$ is diagonalizable within the real space. Due to anti-linearity of $j$ we have $ji= -ij$, which means that if $u$ is an eigenvector of $j$ then $i(u)$ is also an eigenvector, but with an opposite eigenvalue. This implies that the space of states can be decomposed into a direct sum $V= W \oplus i(W)$, where $W\subset V$ is an $n$-dimensional real subspace of the eigenvectors of $j$ with the unit eigenvalue. Equivalently it can be represented as
\begin{eqnarray}
\label{j-general-even} V= \mathbb{C} \otimes_{\mathbb{R}} W, \;\;\; j(\lambda \otimes u)= \lambda^{*} \otimes u,
\end{eqnarray}
and the allowed electronic Hamiltonians are represented by real hermitian matrices that represent operators acting in $W$. We can always choose the basis sets to belong to $W$, so that the basis set transformations that preserve the scalar product are orthogonal, i.e., belong to the orthogonal group ${\rm O}(n)$; therefore, hereafter we refer to this case as orthogonal. Note that such orthogonal transformations are the ones that commute with the real structure $j$.

In the simplest case of $n=2$ we have
\begin{eqnarray}
\label{h-2-surfaces} h(\bm{r})= h_{0}(\bm{r})\sigma_{0}+ h_{x}(\bm{r})\sigma_{x}+ h_{z}(\bm{r})\sigma_{z},
\end{eqnarray}
with $\sigma_{0}$ being the unit $2\times 2$ matrix, so that for the two PES to intersect in a generic (maximal rank) situation, referred to as CSs, two conditions $h_{x}(\bm{r})= h_{z}(\bm{r})= 0$ should be satisfied, so that the CSs have codimension $2$, as well known.

Before we switch to the $j^{2}= -1$ case we consider the case of no time-reversal symmetry (in the presence of magnetic field)
For $n$ PES in the absence of time-reversal symmetry an orthonormal basis set can be chosen up to a unitary transformation, so that we are dealing with the unitary symmetry described by the unitary group ${\rm U}(n)$; therefore, hereafter we refer to this case as unitary. In the case under consideration electronic Hamiltonians are described by just Hermitian operators (matrices) without any further conditions, so that in the simplest $n= 2$ case Eq.~(\ref{h-2-surfaces}) adopts a form
\begin{eqnarray}
\label{h-2-surfaces-u} h(\bm{r})= h_{0}(\bm{r})\sigma_{0}+ h_{x}(\bm{r})\sigma_{x}+ h_{y}(\bm{r})\sigma_{y}+  h_{z}(\bm{r})\sigma_{z},
\end{eqnarray}
and the CSs have codimension $3$.

At this point we turn to the case of odd number of electrons, i.e., half-integer total electron spin, which corresponds to $j^{2}= -1$. Similar to the integer spin case we consider an $n$-dimensional complex vector space of states $V$ and view it as a $2n$-dimensional real vector space equipped with two linear maps $i,j: V\to V$. We further introduce the third linear map $k: V\to V$ by $k= ij$. It is verified in a straightforward way that $i$, $j$, and $k$ anti-commute, and $i^{2}= j^{2}= k^{2}= -1$, as well as $ki = j$ and $jk = i$. This implies that we have a well-defined action of the non-commutative {\it division ring} $\mathbb{H}$ of quaternions on $V$. We reiterate that a quaternion is represented $q= a_{0}+ a_{1}i+ a_{2}j+ a_{3}k$, with $(a_{s}|s=0,\ldots,3)$ being a set of four real numbers; addition and  multiplication of quaternions is defined in an obvious way. As a vector space $\mathbb{H} \cong \mathbb{C}^{2} \cong \mathbb{R}^{4}$. A conjugate $q^{*}$ to $q$ quaternion is naturally defined as
\begin{eqnarray}
\label{q-conjugate} \left(a_{0}+ a_{1}i+ a_{2}j+ a_{3}k\right)^{*}= a_{0}- a_{1}i- a_{2}j- a_{3}k.
\end{eqnarray}
The term division ring means that each nonzero element has an inverse with respect to multiplication, so that sometimes $\mathbb{H}$ is referred to as a non-commutative field.

Although quaternions are non-commutative, their ''field'' property provides a very simple and universal structure of quaternion spaces, e.g., our space of states $V$, it allows for basis sets, and in particular orthonormal basis sets, and a unique decomposition of any state as a linear superposition of the basis set elements with quaternion coefficients. A choice of a basis set allows a representation
\begin{eqnarray}
\label{q-action} V= \mathbb{H} \otimes_{\mathbb{R}} W, \;\;\; q(\lambda \otimes u)= (q\lambda) \otimes u,
\end{eqnarray}
for $u\in W, \; q,\lambda\in \mathbb{H}$, where $W$ is an $m$-dimensional real vector space with $n= 2m$.

Due to the basis set decomposition property transformations between the basis sets are represented by $m\times m$ matrices with quaternion entries. We can, however, further narrow down the class of preferred basis sets. We can apply the analysis of appendix~\ref{sec:Sp-groups} and note that there is a naturally defined action of the group ${\rm Sp}(m)$ on the space $V$ of electronic states. We can also consider a class of {\it real orthonormal} basis sets (see appendix~\ref{sec:Sp-groups} for some details) that are defined as orthonormal basis sets of a special form
\begin{eqnarray}
\label{real-orthonormal} && (e_{1},\ldots, e_{m}, e'_{1},\ldots, e'_{m}) \nonumber \\ && \;\;\; = (e_{1},\ldots, e_{m}, j(e_{1}),\ldots, j(e_{m})).
\end{eqnarray}

Obviously, an invertible linear map $A: V \to V$ belongs to ${\rm Sp}(m)$ if and only if it transfers any real orthonormal basis set to the basis set of the same kind. In physics terms we can say that in the case of time reversal symmetry and $j^{2}= -1$ (odd number of electrons), when we have $n= 2m$ PES, we are dealing with symplectic symmetry, described by the group ${\rm Sp}(m)$; therefore, hereafter we refer to this case as symplectic. Note that since $[j,\hat{H}]= 0$ an adiabatic basis set can be always chosen to be real orthonormal, and all PES are double degenerate, being represented by pairs of adiabatic states $(e_{a}, j(e_{a}))$.

We are now in a position to identify the electronic Hamiltonians $h$, represented by Hermitian operators that commute with $j$. Using the quaternionic representation we find that they are given by $m\times m$ quaternionic matrices with $h_{ba}= h_{ab}^{*}$. In the simplest $m= 2$ case of $n= 2m= 4$ double degenerate PES, and omitting the unit matrix that has nothing to do with the intersections, so that we can deal with traceless matrices, we obtain a $5$-dimensional space of matrices with a basis set to be chosen, e.g., as
\begin{eqnarray}
\label{gamma-matrices} && \gamma_{1}= \left(\begin{array}{cc} 0 & i \\ -i & 0 \end{array}\right) \; \gamma_{2}= \left(\begin{array}{cc} 0 & j \\ -j & 0 \end{array}\right) \; \gamma_{3}= \left(\begin{array}{cc} 0 & k \\ -k & 0 \end{array}\right), \nonumber \\ && \gamma_{4}= \left(\begin{array}{cc} 0 & 1 \\ 1 & 0 \end{array}\right) \; \gamma_{5}= \left(\begin{array}{cc} -1 & 0 \\ 0 & 1 \end{array}\right) \; \gamma_{0}= \left(\begin{array}{cc} 1 & 0 \\ 0 & 1 \end{array}\right)
\end{eqnarray}
A straightforward computation yields
\begin{eqnarray}
\label{gamma-commutation} \gamma_{a}\gamma_{b}+ \gamma_{b}\gamma_{a}= 2\delta_{ab}\gamma_{0}, \;\;\; \gamma_{5}= \gamma_{1}\gamma_{2}\gamma_{3} \gamma_{4},
\end{eqnarray}
with $\gamma_{0}$ being the unit $2\times 2$ quaternionic matrix. By implementing a standard $2 \times 2$ matrix representation of quaternion units
\begin{eqnarray}
\label{quaternion-to-Pauli} 1 \mapsto \sigma_{0}, \;\;\; i \mapsto i \sigma_{x}, \;\;\; j \mapsto i \sigma_{y}, \;\;\; k \mapsto i \sigma_{z},
\end{eqnarray}
we can view the $2\times 2$ quaternionic matrices as $4\times 4$ complex matrices that represent linear operators acting in $V$ in a real orthonormal basis set. Upon substitution of Eq.~(\ref{quaternion-to-Pauli}) into Eq.~(\ref{gamma-matrices}) one can recognize $(\gamma_{a}|a=1,\ldots,4)$ as Euclidean Dirac gamma-matrices, written in the so-called chiral representation, with $\gamma_{5}$ being the product of four Dirac $\gamma$-matrices, so that $(\gamma_{a}| a=1, \ldots, 5)$ represent the five gamma-matrices, associated with the spinors in $5$-dimensional space This implies that an electron Hamiltonian (with the unit matrix omitted) that preserves time-reversal symmetry adopts a form
\begin{eqnarray}
\label{h-2-surfaces-odd} h= \bm{h}\cdot \bm{\gamma}= \sum_{a= 1}^{5}h_{a}\gamma_{a},
\end{eqnarray}
with real coefficients $h_{a}$.

For the Hamiltonians in Eq.~(\ref{h-2-surfaces-odd}) we have two double-degenerate PES with the energies $\varepsilon= \pm \sqrt{(\bm{h}, \bm{h})}$, and the CI (Dirac) point at $\bm{h}= 0$. Since for the conical points, associated with the nuclear configurations, five equations $\bm{h}(\bm{r})= 0$, should be satisfied, the CSs for the time-reversal symmetry with an odd number of electrons have codimension $5$.

\section{Conical Points and Associated Topological Invariants}
\label{sec:conical-topol-inv}

We are now in a position to describe and compare the topological invariants associated with CSs. We start with the unitary case that corresponds to systems of even number of electrons with time-reversal symmetry. Consider a $2$-dimensional vector space of electronic Hamiltonians $h= h_{x}\sigma_{x}+ h_{z}\sigma_{z}$, with the unit matrix that has nothing to do with the PES intersections omitted from Eq.~(\ref{h-2-surfaces}). We have for the electronic energies $\varepsilon= \pm\sqrt{h_{x}^{2}+ h_{z}^{2}}$, so that we have a CS at the origin $h_{x}= h_{z}= 0$. We further surround the origin with a circle $S^{1}$, defined say by $h_{x}^{2}+ h_{z}^{2}= \varepsilon^{2}$. With each point $(h_{x},h_{z})$ of the circle one can associate a $1$-dimensional real space of real eigenstates, say with the higher eigenvalue $\sqrt{h_{x}^{2}+ h_{z}^{2}}$, and further ask a question whether one can identify globally an adiabatic real normalized basis set, i.e., associate with each $1$-dimensional eigenspace a unit length vector in a continuous way. The answer is negative, since upon going over the circle the eigenstate changes the sign. In physics/chemistry literature it is known as the topological Berry phase, which assumes discrete values $0,\pi$. In geometry/topology language one would say that the aforementioned $1$-dimensional bundle has a nontrivial {\it first Stiefel-Whitney class} $w_{1}$ \cite{milnor2016characteristic}, which is binary, rather than integer i.e., resides in $\mathbb{Z}_{2}$, rather than $\mathbb{Z}$, or in other words is represented by a sign $\pm 1$ factor. If we wind a circle around the conical seam, it will be mapped to the space of electronic Hamiltonians by means of Eq.~(\ref{h-2-surfaces}), which will give rise to the topological Berry phase in its conventional sense (gaining a $-1$ factor upon winding around the conical seam). In other words the topological Berry phase in the space of nuclear configurations $\bm{r}$ is completely induced by its counterpart in the space of electronic Hamiltonians, the latter being described above.

In the unitary case of no time-reversal symmetry (e.g., in a famous example of a single spin $1/2$ in a magnetic field) we consider a $3$-dimensional vector space of electronic Hamiltonians $h= h_{x}\sigma_{x}+ h_{y}\sigma_{y}+ h_{z}\sigma_{z}$, with the unit matrix that has nothing to do with the PES intersections omitted from Eq.~(\ref{h-2-surfaces-u}). Similar to the time-reversal case we have the electronic energies $\varepsilon= \pm\sqrt{h_{x}^{2}+ h_{y}^{2}+ h_{z}^{2}}$, so that we have a conical (sometimes also referred to as {\it diabolic}) intersection at the origin $h_{x}= h_{y}= h_{z}= 0$. We further surround the origin with a ($2$-dimensional) sphere $S^{2}$, defined say by $h_{x}^{2}+ h_{y}^{2}+ h_{z}^{2}= \varepsilon^{2}$. With each point $(h_{x}, h_{y}, h_{z})$ of the sphere one can associate a $1$-dimensional complex space of eigenstates, say with the higher eigenvalue 
$\varepsilon$, and further ask a question whether one can identify globally a normalized adiabatic basis set, i.e., associate with each $1$-dimensional eigenspace a unit length vector in a continuous way. The answer is negative again, and this can be rationalized as follows.

Denoting $\bm{h} = \varepsilon (\bm{n}, n_{z})$ and $n_{\pm} = n_{x} \pm in_{y}$, we can recast the eigenvalue problem $(\bm{h} \cdot \bm{\sigma}) \bm{\psi} = \varepsilon \bm{\psi}$, for $\bm{\psi} = (\psi_{1}, \psi_{2})$ in a form
\begin{eqnarray}
\label{eigenvalue-U} (1 - n_{z}) \psi_{1} - n_{-} \psi_{2} &=& 0, \nonumber \\ - n_{+} \psi_{1} + (1 + n_{z}) \psi_{2} &=& 0,
\end{eqnarray}
where the two equations are equivalent. Two equivalent solutions can be naturally identified as
\begin{eqnarray}
\label{eigenvalue-U-Sol} && (\psi_{1}^{-}, \psi_{2}^{-}) = (n_{-}, 1 - n_{z}), \nonumber \\ && (\psi_{1}^{+}, \psi_{2}^{+}) = (1 + n_{z}, n_{+})
\end{eqnarray}
with $\bar{\bm{\psi}}^{\pm} = (1/\sqrt{2 (1 \pm n_{z})}) \, \bm{\psi}_{\pm}$ being the normalized counterparts. The solutions $\bm{\psi}^{\pm}$ turn to zero at the south and north poles of the sphere, respectively, which already provides evidence of an impossibility of building a global adiabatic basis set.

We can extend the aforementioned evidence to a more rigorous argument. To that end we note that the two normalized solutions, both representing a normalized adiabatic state should be connected $\bar{\bm{\psi}}^{+} = g \bar{\bm{\psi}}^{-}$ with $g (\bm{n}, n_{z})$ being a function that admits values in unimodular complex numbers. We can easily see from Eq.~(\ref{eigenvalue-U-Sol}) that $g (\bm{n}, n_{z}) = n_{+} \sqrt{n_{+}^{*} n_{+}}$. Restricting $g$ to any circle that misses both poles, e.g., to equator we obtain the map $g : S^{1} \to {\rm U}(1)$ that is topologically non-trivial, have a nonzero {\it degree} ${\rm deg} \, g = 1$, where the degree can be defined in an integral form
\begin{eqnarray}
\label{define-degree-S-1} {\rm deg} \, g = \frac{1}{2\pi} \int_{S^{1}} g^{-1} \frac{dg}{dx} dx,
\end{eqnarray}
or equivalently as the winding number that measures how many times $g(s)$ winds around the target circle that represents ${\rm U}(1)$, while $s$ winds once around the domain circle $S^{1}$. This implies that an adiabatic basis set $\psi^{-}$ that is defined globally on the southern hemisphere, being recast on the equator in terms of $\psi^{+}$, defined globally on the northern counterpart, using a topologically no-trivial map $g$ may not be contracted on the northern hemisphere, so that a global basis set on the whole sphere does not exist.

Similar to the orthogonal case, there is a topological obstruction to having a global adiabatic basis set, induced by the conical seam, however in the unitary case it is the first {\it Chern class} $c_{1}$ \cite{milnor2016characteristic} that is integer-valued, rather than the binary first Stiefel-Whitney class. To demonstrate that we consider the diagonal components of the non-adiabatic coupling terms, defined with respect to the adiabatic basis sets $\psi^{\pm}$ on the northern and southern hemispheres, and represented by vector potentials/gauge fields $A_{j}^{\pm}$, respectively. By Stokes theorem we have
\begin{eqnarray}
\label{Stokes-low-dim} \int_{S^{1}} A_{j}^{\pm} dx^{j} = \pm \int_{D_{\pm}} dA^{\pm} = \pm \int_{D_{\pm}} \varepsilon^{jk} F_{jk} d^{2}x
\end{eqnarray}
with $F_{jk} = (1/2) (\partial_{j} A_{k} - \partial_{k} A_{j})$ and $\varepsilon^{jk}$ being the vector potential curvature (magnetic field) and Levi-Civita symbol, respectively. The vector potentials are naturally connected by the gauge transformation
\begin{eqnarray}
\label{gauge-transform-i} A_{j}^{+} = A_{j}^{-} + g^{-1} \partial_{j} g,
\end{eqnarray}
Integrating Eq.~(\ref{gauge-transform-i}) over the equator, followed by making use of Eqs.~(\ref{Stokes-low-dim}) and (\ref{define-degree-S-1}) we arrive at
\begin{eqnarray}
\label{first-chern-class} \frac{1}{2 \pi} \int_{S^{2}} \varepsilon^{jk} F_{jk} d^{2}x = {\rm deg} \, g.
\end{eqnarray}
The l.h.s. of Eq.~(\ref{first-chern-class}) is known as an integral representation of the first Chern class $c_{1}$ \cite{milnor2016characteristic}, so that we have $c_{1} = {\rm deg} \, g$, which identifies the first Chern class as the topological invariant, associated with conical intersections in the unitary case.

The symplectic case of time-reversal symmetry for systems with odd number of electrons is treated very similar to the unitary case: in the relevant situation of two Kramers doublets we surround a conical point $\bm{h}= 0$ with a $4$-dimensional sphere $S^{4}$, defined, say, by a condition $(\bm{h}, \bm{h})= \varepsilon^{2}$. With each point $\bm{h}$ of the sphere one can associate a $2$-dimensional complex subspace of double-degenerate eigenstates, say with the higher eigenvalue, in the $4$-dimensional complex state of electronic states under consideration, which, according to the quaternionic approach, presented in section~\ref{sec:kramers-symmetry}, is equivalent to associating with each point a $1$-dimensional quaternion vector subspace of the $2$-dimensional quaternion space of electronic states. We further ask a question whether one can identify a global real orthonormal adiabatic basis set, i.e., associate with each $2$-dimensional eigenspace a real orthonormal basis set, i.e. a pair $(\bm{e},j(\bm{e}))$ with $|\bm{e}|= 1$, in a continuous way, which is equivalent to identifying a quaternion vector function $\bm{\psi} (\bm{h})$, with $\bm{\psi} = (\psi_{1}, \psi_{2})$, that satisfies the eigenvalue problem. The answer is negative again, and this can be rationalized exactly in the same way as for the unitary case.

Indeed, denoting $\bm{h} = \varepsilon (\bm{n}, n_{z})$ and $n_{\pm} = n_{4} \mp i n_{1} \mp j n_{1} \mp k n_{3}$, the eigenvalue problem $(\bm {h} \cdot \bm{\gamma}) \bm{\psi} = \varepsilon \bm{\psi}$, with the gamma-matrices given by Eq.~(\ref{gamma-matrices}) adopts the form of Eq.~(\ref{eigenvalue-U}) and naturally has the same solution as in the unitary case, given by Eq.~(\ref{eigenvalue-U-Sol}), with the only difference that $n_{\pm}$ are quaternions, rather than complex numbers, and $(\bm{n}, n_{z})$ resides in the $4$-dimensional sphere $S^{4}$, rather than its $2$-dimensional counterpart $S^{2}$. In particular, the solutions $\bm{\psi}^{\pm}$ have zeros at the south and north pole of $S^{4}$, while their normalized counterparts are connected
\begin{eqnarray}
\label{sections-connect} \bar{\bm{\psi}}^{+} = g \bar{\bm{\psi}}^{-}, \;\;\; g(\bm{n}, n_{z}) = \frac{n_{+}}{\sqrt{n_{+}^{*} n_{+}}}
\end{eqnarray}
via the function $g$ that admits values in unit length quaternions, the latter forming the group ${\rm Sp} (1)$, which by construction, as a space, forms a $3$-dimensional sphere $S^{3}$. Note that Eq.~(\ref{quaternion-to-Pauli}) establishes an isomorphism ${\rm Sp} (1) \cong {\rm SU} (2)$, so that being restricted to $S^{3} \subset S^{4}$, say, by fixing the value of $n_{z} \ne \pm 1$, e.g., to the equator for $n_{z} = 0$, we obtain a map $g : S^{3} \to {\rm SU} (2)$, which is, in complete analogy with the unitary case is topologically non-trivial, which can be established by generalizing the notion of the degree of a map $g : S^{n} \to S^{n}$ from the case $n=1$, considered earlier, and given by the winding number, to the case of any natural $n$. To that end we note that the winding number of $g : S^{1} \to S^{1}$ can be measured by performing weighted counting of how many times $g(s)$ crosses some arbitrarily chosen reference point in the target $S^{1}$, while $s$ winds once along the domain $S^{1}$, with the weights represented by $\pm 1$ sign factors depending on the direction in which $g(s)$ goes through the reference point. The described procedure can be easily generalized to the arbitrary dimension case by looking at the generically finite set $g^{-1} (\{s_{0}\})$ of preimages of some arbitrary chosen reference point $s_{0} \in S^{n}$ and counting the preimages, weighting them with sign factors, given by the signs of the Jacobian of $g$ at the corresponding points. More formally we define ${\rm deg} \, g = \sum_{s \in g^{-1}(s_{0})} {\rm sgn} (\det (\partial  g(s) / \partial s))$. In complete analogy with the unitary case, we see that the map $g : S^{3} \to S^{3}$, defined above, is one-to-one, and therefore, having a non-zero degree ${\rm deg} \, g = 1$, is topologically non-trivial, so that all arguments on the topologically nontrivial structure, introduced by a CS, presented above for the unitary case, work in the symplectic situation in the exactly same way.

Similar to the unitary case the degree of $g$ can be related to the Chern class, however, for the symplectic situation it is the second Chern class $c_{2}$ \cite{milnor2016characteristic}.
To see that we note that, in complete analogy with the unitary case, there is an integral representation for the degree of our map $g : S^{3} \to {\rm SU} (2)$
\begin{eqnarray}
\label{define-degree-S-3} {\rm deg} \, g = \frac{1}{24 \pi^{2}} \int_{S^{3}} {\rm Tr} (g^{-1}dg \wedge g^{-1}dg \wedge g^{-1}dg),
\end{eqnarray}
rationalized by the fact (see appendix~\ref{sec:diff-forms-Chern-cl} for a more formal argument) that, up to a normalization constant, the integrand is given by the Jacobian of $g$. Therefore, the original integral over the domain of $g$ can be interpreted as the integral of a constant function (whose value is determined by the aforementioned normalization constant) over the target space of $g$, multiplied by an integer factor that accounts for the multiplicity of the preimages of points in the target space. Recalling the definition of the map degree presented above, it becomes intuitively clear that this factor is given by ${\rm deg} \, g$. Using a similar to the unitary, still more technically involved approach (see appendix~\ref{sec:diff-forms-Chern-cl} for some details), and treating $g$ as a gauge transformation of the diagonal non-adiabatic coupling terms, the latter being considered as a non-abelian (Yang-Mills) ${\rm SU} (2)$ gauge field, described by the matrix vector potential $A_{j} = -i\sum_{a=1}^{3} A_{j}^{a} \sigma_{a}$  Eq.~(\ref{define-degree-S-3}) can be recast in a form
\begin{eqnarray}
\label{second-chern-class} \frac{1}{8\pi^{2}} \int_{S^{4}} {\rm Tr} (F \wedge F) = {\rm deg} \, g.
\end{eqnarray}
with $F = F_{jk} dx^{j} \wedge dx^{k}$, where $F_{jk} = (1/2) (\partial_{j} A_{k} - \partial_{k} A_{j} + [A_{j}, A_{k}])$ is the non-abelian curvature. One recognizes the l.h.s. as a standard integral representation of the second Chern class $c_{2}$ \cite{milnor2016characteristic}, identifying it as the topological invariant, associated with conical intersections in the symplectic case. Some details of a derivation of Eq.~(\ref{second-chern-class}) from Eq.~(\ref{define-degree-S-3}), more formal rationalization of the latter, explanation why Eq.~(\ref{second-chern-class}) reproduces the second Chern class, as well as necessary facts and concepts, associated with differential forms, including wedge products and Stokes theorem, involved in the aforementioned derivations, are presented in appendix~\ref{sec:diff-forms-Chern-cl}.

We conclude this section with noting that as opposed to the orthogonal, in the unitary and symplectic cases the proper adiabatic states are defined up to a continuous degree of freedom, which sits in ${\rm U}(1)$ and ${\rm SU} (2)$, respectively, giving rise to diagonal components of the nonadiabatic coupling terms. The corresponding vector fields $A_{j}$ are geometrically non-trivial, i.e., they have non-zero curvature $F_{jk}$, so that in the unitary case the effect of geometric (i.e., path-dependent) Berry phase takes place. A similar effect occurs in the symplectic case, where instead of the phase, as an element of ${\rm U} (1)$, we have an element of ${\rm SU} (2)$, hereafter referred to as {\it non-abelian Berry phase} \cite{Pacher1989}. The latter will be discussed in some detail in section~\ref{sec:BO-semicl-half-int}.

\section{Born-Oppenheimer Approximation for Half-Integer Spin Case, Semiclassical Propagation, and Non-Abelian Berry Phase}
\label{sec:BO-semicl-half-int}

The easiest way to rationalize semiclassical adiabatic dynamics for systems with time-reversal symmetry and odd number of electrons (symplectic case) is no bring in the partial path integral representation with matrix action, introduced, e.g., in~\cite{Piryatinski2005}, where the path integration is performed over the nuclear position variables $\bm{r}$, whereas the electronic counterparts are treated explicitly. Being focused on the case of two (both double-degenerate) potential surfaces, and following~\cite{Piryatinski2005}, we represent the Hamiltonian in a form
\begin{eqnarray}
\label{BO-generalized} H = - \sum_{j=1}^{d} \frac{\hbar^{2} \nabla_{j}^{2}}{2 m_{j}} + \sum_{\alpha=0}^{5} {\cal U}_{\alpha}(\bm{r}) \gamma_{a},
\end{eqnarray}
with $\nabla_{j} = \partial/\partial r_{j} - A_{j}$ being the ''long'' gauge-invariant derivatives. The Hamiltonian in Eq.~(\ref{BO-generalized}) can be viewed as a generalized $2$-state Born-Oppenheimer (BO) approximation, with two double-degenerate PES. It treats adequately intersections of the two chosen PES, and requires only the rest of PES to be separated energetically, so that nonadiabatic coupling to them can be neglected. It is obtained by projecting the original Hamiltonian to the electronic subspace spanned onto the adiabatic states, which results in the standard expressions $A_{j}^{ab} = \langle \psi_{a}(\bm{r}) | \partial \psi_{b}(\bm{r})/\partial r_{j} \rangle$, where $\psi_{a} (\bm{r})$, with $a = 1, \ldots, 4$ being some position-dependent orthonormal real (in the sense of section~\ref{sec:kramers-symmetry}) basis set in the space of electronic sates.

Assuming we are far away from CSs, we further apply the complete BO approximation, which boils down to choosing an adiabatic basis set and neglecting the block off-diagonal components of $A_{j}^{ab}$, i.e. the ones with $a$ and $b$ belonging to different adiabatic surfaces, making evolution on both surfaces independent of each other. The corresponding BO Hamiltonians have a form
\begin{eqnarray}
\label{BO-symplectic} H = - \sum_{j=1}^{d} \frac{\hbar^{2} \nabla_{j}^{2}}{2 m_{j}} + {\cal U}(\bm{r}),
\end{eqnarray}
with
\begin{eqnarray}
\label{BO-symplectic-2} {\cal U}(\bm{r}) = {\cal U}_{0}(\bm{r}) \pm |\bm{{\cal U}(\bm{r})}|
\end{eqnarray}
being the adiabatic energies, whereas the diagonal, in the aforementioned sense components $A_{j}$ are represented by $2 \times 2$ matrices $A_{j}^{ab}$, defined with respect to an orthonormal real basis set $\psi_{a}(\bm{r})$, with $\psi_{2} = j \psi_{1}$, and therefore,
\begin{eqnarray}
\label{BO-symplectic-3} A_{j}(\bm{r}) = -i \sum_{\mu=1}^{3} A_{j}^{\mu}(\bm{r}) \sigma_{\mu}.
\end{eqnarray}
The difference between the adiabatic evolution in the orthogonal and symplectic cases is that in the latter the wavefunction has a $2$-component vector character and there is a non-abelian matrix gauge field that elongates the spatial derivatives.

Applying the path-integral representation to the evolution operator, associated with the adiabatic Hamiltonian [Eq.~(\ref{BO-symplectic})] in a way, described in the beginning of this section, we obtain
\begin{eqnarray}
\label{G-path-int} \hat{G} (\bm{r}'', \bm{r}'; t) = \int_{\bm{x}(0)=\bm{r}'}^{\bm{x}(t)=\bm{r}''} {\cal D}\bm{x} \exp \left(\frac{i}{\hbar} S(\bm{x})\right) \hat{U}(\bm{x})
\end{eqnarray}
with
\begin{eqnarray}
\label{G-path-int-texp} && S(\bm{x}) = \int_{0}^{t} d\tau \left(\frac{m \dot{\bm{x}}^{2}(\tau)}{2} - {\cal U}_{0}(\bm{x}(\tau))\right), \nonumber \\ && \hat{U}(\bm{x}) = T \exp \left(\int_{\bm{x}} d\bm{r} \cdot \bm{A}\right).
\end{eqnarray}

The semiclassical adiabatic propagator is obtained by neglecting the trajectory fluctuations around the classical counterpart in computing $\hat{U}(\bm{x})$, so that the path integral represent just the standard adiabatic propagator, followed by applying the van Vleck semiclassical approximation to the latter, resulting in
\begin{eqnarray}
\label{G-semiclass} \hat{G} (\bm{r}'', \bm{r}'; t) = G_{0} (\bm{r}'', \bm{r}'; t) \hat{U} (\bm{x}_{\rm cl} (\bm{r}'', \bm{r}'; t))
\end{eqnarray}
with $G_{0}$ denoting the van Vleck semiclassical propagator,
so that Eq.~(\ref{G-semiclass}) solves the problem of adiabatic dynamics in the semiclassical approximation. Semiclassical evolution near CSs, where the adiabatic approximation breaks down is considered in section~\ref{sec:conical-semicl-half-int}.

The vector character of a Kramers doublet, considered in this section is naturally described in terms of the wavepacket polarization $\zeta(\bm{r})$ defined by the conditions $\Psi(\bm{r}) = |\Psi(\bm{r})| \zeta(\bm{r})$ and $|\zeta(\bm{r})| = 1$, so that the polarization is represented by a nuclear position dependent unit vector in the $2$-dimensional complex vector space of electronic states of a Kramers doublet, so that the polarization $\zeta(\bm{r}) \in S^{3}$ resides in a $3$-dimensional sphere. Obviously the second (matrix) factor in the r.h.s. of Eq.~(\ref{G-semiclass}) affects the polarization only, keeping $|\Psi(\bm{r})|$ unchanged. However, the first (scalar) factor that represents the standard Van Vleck propagator, also affects the polarization, e.g., due to the phase factor $e^{i\hbar^{-1} S_{\rm cl}}$ that originates from the classical action. Still the evolution of polarization dynamics can be completely decoupled from the scalar Van Vleck evolution via introducing the reduced polarization $\bar{\zeta}(\bm{r})$ by considering two values of polarization $\zeta$ and $\zeta'$, represented by two unit $2$-dimensional complex vectors, the same, if the latter differ by a unimodular factor. The reduced space of the described above equivalence classes is represented by the complex projective line $\mathbb{C}P^{1}$, the latter being topologically equivalent to the $2$-sphere $S^{2}$. The reduction map $S^{3} \to S^{2}$ that maps the polarization to its reduced counterpart is known in topology as a Hopf map. The aforementioned unimodular factor can be absorbed by the scalar part of the nuclear wavefunction, so that the latter can be represented by a complex-valued scalar wavefunction and reduced polarization $\bar{\zeta}$, instead of the polarization $\zeta$ and a real ``wavefunction'' $|\Psi|$, so that within the new (reduced polarization) representation picture, the scalar (Van Vleck) and polarization evolution are completely decoupled.

\begin{figure}[!h]
	\includegraphics[]{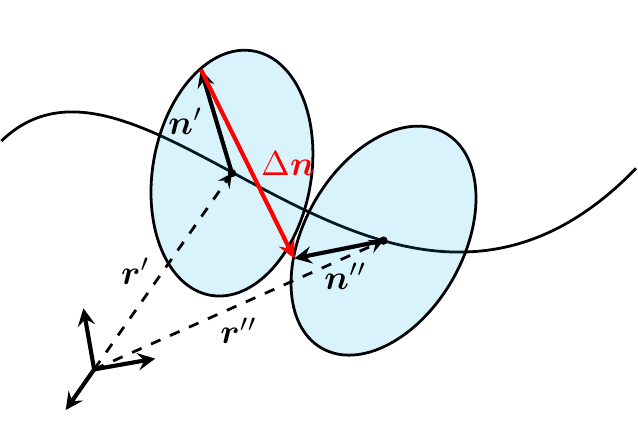}
	\caption{Geometric nature of wavepacket polarization evolution, described by parallel transport. The wavepacket polarizations $\bm{n}'$ and $\bm{n}''$ at different times belong to different subspaces. The new polarization value $\bm{n}''$, interpreted as a result of parallel transport over an infinitesimal time period, is uniquely determined by two conditions: $\bm{n}''$ should be normalized, and the polarization change $\Delta \bm{n}$ should be orthogonal to its initial value $\bm{n}'$.}
	\label{fig:spin-symplectic-polarization}
\end{figure}

It follows immediately from Eq.~(\ref{G-semiclass}) that semiclassical evolution of the wavepacket (reduced) polarization is of completely geometric nature, and is related to multiple phenomena, which, in particular include adiabatic propagation of a spin in time-dependent magnetic field, rotating cats/astronauts, stochastic current, generated by adiabatic driving, and are often referred to as Berry phase phenomena. Indeed, the geometrical meaning of the vector potential/gauge field that represents the diagonal component of the nonadiabatic coupling terms is that it determines the parallel transport of the electronic state along a trajectory, as illustrated in Fig. \ref{fig:spin-symplectic-polarization}. This is why in differential geometry it is referred to as a {\it connection}. The same connection appears in a different setting when the electronic Hamiltonian depends not on additional variables, in our case nuclear coordinates, but rather just on time. In the unitary case this would be a problem of a spin in a time-dependent magnetic field; in this case $\hat{U}(C)$, with $C$ being a path in the $3$-dimensional space of electronic Hamiltonians $\bm{h}$, belongs to ${\rm U}(1)$ and for a closed path (loop) reproduces exactly the celebrated Berry phase. In the symplectic case the phase becomes non-abelian, i.e., it belongs to ${\rm SU}(2)$, as outlined in~\cite{Pacher1989}.

\section{Semiclassical Theory for Nuclear Wavepacket Propagation Through a Conical Seam}
\label{sec:conical-semicl-half-int}

In this section we obtain explicit asymptotically exact expressions for the evolution of nuclear wavepackets in the presence of CSs in the semiclassical regime for all three situations, with focus on the symplectic case that corresponds to the half-integer spin. Compared to the integer-spin counterpart the half-integer situation is treated in a similar way, it is just technically more involved. As outlined in~\cite{Piryatinski2005} for the ''standard'' integer spin case, as long as the wavepacket is not close to a CS, i.e., outside of the conical scattering region, it is moving adiabatically, which means that in the semiclassical regime a standard Van Vleck semiclassical propagator can be applied for asymptotically exact description of the system evolution (we reiterate that the Van Vleck propagator approach is equivalent to the Gaussian Ansatz for wavepackets). When the wavepacket passes through the CS, ballistic approximation is valid in the semiclassical regime. The described approach in fact works due to the overlap of the adiabatic and ballistic regions, as clearly demonstrated in~\cite{Piryatinski2005}. The semiclassical approach for the adiabatic region has been extended to the half-integer spin case in section~\ref{sec:BO-semicl-half-int}, including effects of the non-abelian Berry phase.

Since the goal of~\cite{Piryatinski2005} was to extend the Van Vleck semiclassical propagator to the case of the presence of CSs, evolution in the scattering region was described on the level of the ballistic propagator, which was obtained by bringing in the path-integral approach with matrix contribution to the action, followed by neglecting the fluctuation of the nuclear trajectory in computing the time-ordered exponential, associated with the matrix component of the action. Of course the wavepacket evolution in the vicinity of a CS can be readily obtained by applying the ballistic propagator to the incoming wavepacket, however, in this manuscript we will derive explicit expressions for the wavepacket evolution directly from the dynamical Schr\"odinger equation. The advantages of this way include simplicity of the derivation, bypassing additional integration involved in applying the propagator to the incoming wavepacket, as well as relative easiness in connecting the ballistic and adiabatic solutions in the overlap region.

The ballistic approximation starts with switching to a diabatic basis set (the use of the indefinite article is important), defined by a condition ${\bm{A}} (\bm{r}_{0}) = 0$, with the point $\bm{r}_{0}$, where the wavepacket, whose size scales $\sim \sqrt{\hbar}$, crosses the conical seam, being well defined in the semiclassical $\hbar \to 0$ limit, followed by introducing the time-dependent wavepacket position $\bm{R}(t) = \bm{R}_{0} + \bm{v} (t - t_{0})$, and representing the system wavefunction in a form
\begin{eqnarray}
\label{Psi-ballistic} \Psi (\bm{r}, t) = \exp (i\hbar^{-1} \bm{p} \cdot (\bm{r} - \bm{R}(t))) \bar{\Psi} (\bm{r} - \bm{R}(t), t),
\end{eqnarray}
with $\bm{p} = m \bm v$ and $\bm{v}$ being the wavepacket momentum and velocity, respectively. Upon substitution of Eq.~(\ref{Psi-ballistic}) into the dynamical Schr\"odinger equation we obtain
\begin{eqnarray}
\label{SE-ballistic} i\hbar \frac{\partial \bar{\Psi} (\bm{r}, t)}{\partial t} = (H_{\rm B}(t) + H_{1}) \bar{\Psi} (\bm{r}, t)
\end{eqnarray}
with
\begin{eqnarray}
\label{SE-ballistic-H} H_{\rm B}(t) = -\frac{\bm{p}^{2}}{2m} + \hat{h} (\bm{r}_{\rm L}(\bm{r}, t)), \;\;\; H_{1} = -\frac{\hbar^{2} \partial^{2}}{2m},
\end{eqnarray}
$\bm{r}_{\rm L}(\bm{r}, t) = \bm{R}(t) + \bm{r}$, $\hat{h}(\bm{r}_{\rm L}) = \bm{h}(\bm{r}_{\rm L}) \cdot \bm{\gamma}$, and $\partial^{2}$ being the Laplace operator. The ballistic approximation boils down to neglecting the $H_{1}$ term in the r.h.s. of Eq.~(\ref{SE-ballistic}) turning the PDE [Eq.~(\ref{SE-ballistic})] into a family of ODE
parameterized by $\bm{r}$, whose solutions can be explicitly represented in terms of time-ordered exponentials, resulting in:
\begin{eqnarray}
\label{solution-ballistic}  \bar{\Psi} (\bm{r}, t) &=& e^{iS_{\rm B} / \hbar} T \exp \left( - \frac{i}{\hbar v^{2}} \int_{C} \hat{h} (\bm{r}') \bm{v} \cdot d\bm{r}'\right) \nonumber \\ && \;\;\;\; \times \bar{\Psi} (\bm{r}, t_{0}),
\end{eqnarray}
with $C$ and
\begin{eqnarray}
\label{solution-ballistic-2} S_{\rm B} = \frac{m \bm{v}^{2} (t - t_{0})}{2}
\end{eqnarray}
being the straight (ballistic) path that connects $r_{\rm L} (\bm{r}, t_{0})$ to $r_{\rm L} (\bm{r}, t)$, and the ballistic action, respectively.

\begin{figure}[!h]
	\includegraphics[width=0.45\textwidth]{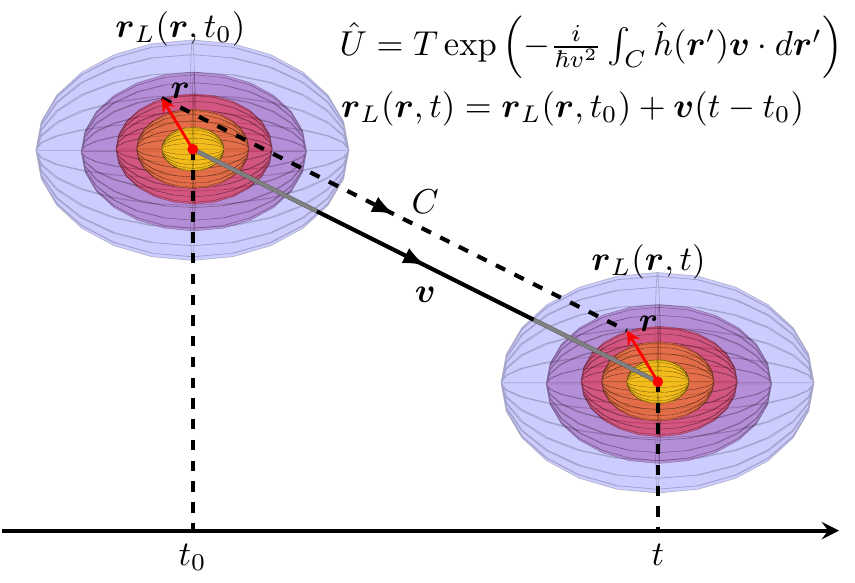}
	\caption{Illustration of ballistic wavepacket propagation, using the laboratory frame, in which the wavepacket moves with a constant velocity $\bm{v}$, i.e., ballistically. The molecular configuration vector $\bm{r}$ in the proper frame, associated with a moving wavepacket, shown as red, stays unchanged, while its laboratory frame counterpart changes from $\bm{r}_{L}(\bm{r}, t_{0})$ to $\bm{r}_{L}(\bm{r}, t)$. The dashed line represents the integration path $C$ in Eq.~ (\ref{solution-ballistic}).}
	\label{fig:spin-ballistic-propagator}
\end{figure}

An explicit expression for the evolution in the ballistic approximation [Eq.~(\ref{solution-ballistic})] has a very simple and natural interpretation, namely there are two factors that affect the evolution: (i) the wavepacket is moving ballistically, i.e., with a constant velocity $\bm{v}$, and (ii) the (vector) value of the wavefunction for any position $\bm{r}$ in the moving frame is evolving according to the value $\hat{h} (\bm{r}_{\rm L} (\bm{r}, t))$ of the matrix Hamiltonian at the corresponding point in the laboratory frame. The aforementioned interpretation is illustrated in Fig. \ref{fig:spin-ballistic-propagator}.

We will apply the ballistic approximation to the region around the conical seam, where the position dependence $\hat{h} (\bm{x})$ can be linearized, We further note that the expression for ballistic propagation [Eq.~(\ref{solution-ballistic})] is valid for short enough times for any matrix Hamiltonian $\hat{h} (\bm{x})$ and a configuration space of any dimension. In particular, the aforementioned expression is capable of handling all three, namely the orthogonal, unitary, and symplectic, cases in the configuration space of arbitrary dimension. On the other hand, during the ballistic process of the wavepacket crossing a CS, nothing happens to the wavepacket shape along the CS, with all changes occurring in the transverse directions. Therefore, for the sake of presentation clarity/simplicity, and to avoid notational clutter we will set the configuration space dimension to $d=2$, $d=3$, and $d=5$, for the orthogonal, unitary, and symplectic cases, respectively, postponing a very simple discussion of a truly straightforward extension to the arbitrary dimension case to
section~\ref{sec:toplology-scatrd}. In all three cases, for the aforementioned dimensions, the CS is represented by a single point, located at the coordinate origin. The derivations, as well as the final expressions, become most compact upon implementing an appropriate coordinate system $\bm{r} = (\bm{x}, z)$ in the configuration space and an appropriate basis set in the relevant subspace of electronic states.

We start with the simplest orthogonal case, in particular setting $d=2$. We first linearize the dependence $\hat{h} (\bm{x})$. We then choose the direction $\bm{e}_{z}$ of the $z$-axis in the wavepacket velocity direction, and further rotate the basis set in the $2$-dimensional electronic space to achieve $\hat{h} (\bm{e}_{z}) = f\sigma_{z}$ for some $f$. We further identify the direction $\bm{e}_{x}$ of the $x$-axis by imposing the condition $\hat{h} (\bm{e}_{x}) = f\sigma_{x}$, to arrive at
\begin{eqnarray}
\label{h-lin-orthog} \hat{h} (x, z) = f (x \sigma_{x} + z \sigma_{z}),
\end{eqnarray}
with $f$ being a (scalar) force constant. Note that it is due to the coordinate/basis choices, described above, we were able to replace a $2 \times 2$ force constant matrix with a single scalar counterpart.

In the unitary case we choose $\bm{e}_{z}$ and rotate the basis set in exactly the same way as in the orthogonal situation, achieving $\hat{h} (\bm{e}_{z}) = f\sigma_{z}$, whereas $\bm{e}_{x}$ and $\bm{e}_{y}$ are identified in a similar way from the conditions $\hat{h} (\bm{e}_{x}) = f\sigma_{x}$ and $\hat{h} (\bm{e}_{y}) = f\sigma_{y}$, resulting in
\begin{eqnarray}
\label{h-lin-unitary} \hat{h} (\bm{x}, z) = f (\bm{x} \cdot \bm{\sigma} + z \sigma_{z}),
\end{eqnarray}
with $\bm{x} = (x, y)$ and $\bm{\sigma} = (\sigma_{x}, \sigma_{y})$.

For the symplectic case, in a similar way, we choose $\bm{e}_{5}$ to be along the wavepacket velocity and further achieve $\hat{h} (\bm{e}_{5}) = f\gamma_{5}$ via the electronic space basis set choice, and further identify $\bm{e}_{j}$ from the conditions $\hat{h} (\bm{e}_{j}) = f\gamma_{j}$, for $j = 1, 2, 3, 4$. This results in
\begin{eqnarray}
\label{h-lin-sympl} \hat{h} (\bm{x}, z) = f (\bm{x} \cdot \bm{\gamma} + z \sigma_{z}),
\end{eqnarray}
with $\bm{x} = (x_{1}, x_{2}, x_{3}, x_{4})$ and $\bm{\gamma} = (\gamma_{1}, \gamma_{2}, \gamma_{3}, \gamma_{4})$. Note that the unitary case [Eq.~(\ref{h-lin-unitary})] can be represented in the form of Eq.~(\ref{h-lin-sympl}) by setting $\bm{\gamma} = (\sigma_{x}, \sigma_{y})$ and $\gamma_{5} = \sigma_{z}$.

Following~\cite{Piryatinski2005} we introduce the scattering $r_{\rm s}$ and ballistic $r_{\rm B}$ length scales
\begin{eqnarray}
\label{define-r-s-b} r_{\rm s} = \sqrt{{\hbar v}/{f}}, \;\;\; r_{\rm B} = \left({m \hbar v^{3}}/{f^{2}}\right)^{1/3},
\end{eqnarray}
so that the matching region, where both the ballistic and adiabatic approximation are valid is defined by
\begin{eqnarray}
\label{match-region} r_{\rm s} \ll r \ll  r_{\rm B},
\end{eqnarray}
and the overlap $r_{\rm s} \ll r_{\rm B}$ of the ballistic and adiabatic regions is provided by the condition $g_{\rm s} \ll 1$, with the dimensionless parameter that controls applicability of our semiclassical approach given by
\begin{eqnarray}
\label{define-g-s} g_{\rm s} = \sqrt{f \hbar / m^{2} v^{3}}, \;\;\; r_{\rm s} /  r_{\rm B} = g_{\rm s}^{1/3}.
\end{eqnarray}
for the orthogonal case, we further introduce the dimensionless parameter $l$ that parameterizes ballistic trajectories and the dimensionless impact parameter $\alpha$
\begin{eqnarray}
\label{define-l-alpha} l = (\sqrt{2}/r_{\rm s}) (z_{0} + vt), \;\;\; \alpha = \sqrt{2} x / r_{\rm s},
\end{eqnarray}
so that the time-ordered exponential in Eq.~(\ref{solution-ballistic}) is obtained by solving a linear ODE
\begin{eqnarray}
\label{t-exp-equation} i \frac{d}{dl} \Psi(l) = \frac{1}{2} (\alpha\sigma_{x} + l\sigma_{z}) \Psi(l).
\end{eqnarray}

The time-ordered exponential in Eq.~(\ref{solution-ballistic}) is therefore given by the evolution operator, associated with Eq.~(\ref{t-exp-equation}),
\begin{eqnarray}
\label{evolution-LZ} \hat{U} (l_{2}, l_{1}) =
\left(\begin{array}{cc}
u_{\rm d} (l_{2}, l_{1}) & u_{\rm a} (l_{2}, l_{1}) \\
-u_{\rm a}^{*} (l_{2}, l_{1}) & u_{\rm d}^{*} (l_{2}, l_{1})
\end{array}\right)
\end{eqnarray}
which, in the relevant for us limit $l_{1} \to -\infty$ and $l_{2} \to \infty$, is given by the scattering matrix of the celebrated Landau-Zener (LZ) problem
\begin{eqnarray}
\label{scattering-LZ} && u_{\rm d} = \sqrt{P_{\rm d}} e^{-i \Phi_{\rm d}}, \;\;\; u_{\rm a} = \sqrt{1- P_{\rm d}} e^{-i \Phi_{\rm a}}, \nonumber \\ && \Phi_{\rm d} = \Phi_{2} - \Phi_{1} \;\;\; \Phi_{\rm a} = \Phi_{2} + \Phi_{1} - \delta (\alpha),
\end{eqnarray}
with $P_{\rm d} (\alpha)= \exp (-\pi\alpha^{2}/2)$, $\Phi_{j} = (l_{j}^{2} + \alpha^{2}\ln|l_{j}|)/4 = \Phi (l_{j}; \alpha)$ are the LZ probability to stay on a diabatic level and the adiabatic phases, respectively with $j = 1$ and $j = 2$ corresponding to the initial and final points of a ballistic trajectory. The nonadiabatic phase shift $\delta(\alpha) = \pi/4 - {\rm arg} \, \Gamma (-i\alpha^{2}/4)$ is expressed in terms of the Euler gamma function $\Gamma (z)$.

The expressions, provided by Eqs.~(\ref{evolution-LZ}) and (\ref{scattering-LZ}), being substituted into Eq.~(\ref{solution-ballistic}) fully describe the asymptotically exact semiclassical scattering of a wavepacket on a conical scheme for the orthogonal case. In order to apply them to the unitary and symplectic cases in an almost straightforward way we recast them in a form
\begin{eqnarray}
\label{scattering-LZ-gen} \hat{U} &=& \sqrt{P_{\rm d}} (\gamma_{0} \cos\Phi_{\rm d} - i\gamma_{5} \sin\Phi_{\rm d}) \nonumber \\ &+& \sqrt{1 - P_{\rm d}} (\gamma_{5} \gamma \cos\Phi_{\rm a} - i\gamma \sin\Phi_{\rm a}),
\end{eqnarray}
with $\gamma_{0} = \sigma_{0}$, $\gamma = \sigma_{x}$, and $\gamma_{5} = \sigma_{z}$. Using the introduced notation Eq.~(\ref{t-exp-equation}) is naturally represented in a form
\begin{eqnarray}
\label{t-exp-equation-gen} i \frac{d}{dl} \Psi(l) = \frac{1}{2} (\alpha \gamma + l \gamma_{5}) \Psi(l).
\end{eqnarray}

The key observation on the way of extending our expressions to the unitary and symplectic cases is that both Eq.~(\ref{t-exp-equation-gen}) and the associated evolution operator [Eq.~(\ref{scattering-LZ-gen})] are expressed in terms of an algebra, generated by $\gamma_{0}$, $\gamma$, and $\gamma_{5}$ with the relations $\gamma^{2} = \gamma_{5}^{2} = \gamma_{0}$, $\gamma_{5} \gamma = - \gamma \gamma_{5}$ (anticommute), and $\gamma_{0}$ being the unit. Therefore, for any matrices with the described above relations the evolution operator, associated with Eq.~(\ref{t-exp-equation-gen}), is given by Eq.~(\ref{scattering-LZ-gen}).

In the unitary and symplectic cases, when the position is described by $(\bm{x}, z)$, the impact parameter $\bm{x}$, associated with a ballistic trajectory is of vector nature, and is naturally represented as $\bm{x} = x \bm{n}$, with $\bm{n}$ being a unit vector, so that $x$ can be interpreted as a scalar impact parameter. For the unitary case, defining $\gamma = \bm{n} \cdot \bm{\gamma}$ with, as described above $\gamma_{0} = \sigma_{0}$, $\bm{\gamma} = (\sigma_{x}, \sigma_{y})$, and $\gamma_{5} = \sigma_{z}$, we find that the equation that describes the relevant time-ordered exponential is given by Eq.~(\ref{t-exp-equation-gen}), which immediately implies that the associated evolution operator $\hat{U}$ is given by Eq.~(\ref{scattering-LZ-gen}) with the described above values of the $\gamma$-matrices, so that after some straightforward algebra we arrive at
\begin{eqnarray}
\label{scattering-LZ-U} \hat{U} =
\left(\begin{array}{cc}
\sqrt{P_{\rm d}} e^{-i \Phi_{d}} & \sqrt{1 - P_{\rm d}} e^{-i \Phi_{a}} n_{+} \\
-\sqrt{1 - P_{\rm d}} e^{i \Phi_{a}} n_{-} & \sqrt{P_{\rm d}} e^{i \Phi_{d}}
\end{array}\right)
\end{eqnarray}
with $n_{\pm} = n_{x} \pm i n_{y}$.

The symplectic case is treated exactly in the same way setting $\gamma = \bm{n} \cdot \bm{\gamma}$, with $\gamma_{0}$, $\bm{\gamma} = (\gamma_{1}, \gamma_{2}, \gamma_{3}, \gamma_{4})$, and $\gamma_{5}$ given by Eq.~(\ref{gamma-matrices}). Using a standard matrix representation of the quaternionic units
in terms of the Pauli matrices [Eq.~(\ref{quaternion-to-Pauli})], we obtain upon its substitution into Eq.~(\ref{scattering-LZ-gen}), followed by straightforward algebra
\begin{eqnarray}
\label{scattering-LZ-Sp} \hat{U} =
\left(\begin{array}{cc}
\sqrt{P_{\rm d}} e^{-i \Phi_{d}} \sigma_{0} & \sqrt{1 - P_{\rm d}} e^{-i \Phi_{a}} u(\bm{n}) \\
-\sqrt{1 - P_{\rm d}} e^{i\Phi_{a}} u^{\dagger}(\bm{n})  & \sqrt{P_{\rm d}} e^{i \Phi_{d}} \sigma_{0}
\end{array}\right)
\end{eqnarray}
with $\bm{n} = (\bm{\eta}, n_{4})$, so that $\bm{\eta}^{2} + n_{4}^{2} = 1$ and
\begin{eqnarray}
\label{scattering-LZ-Sp-2} u (\bm{n}) = n_{4} \sigma_{0} + i \bm{\eta} \cdot \bm{\sigma}
\end{eqnarray}
being a direction dependent unitary matrix. Note that Eq.~(\ref{scattering-LZ-Sp-2}) provides a standard global parameterizations of the unitary group, in particular establishing an isomorphism $S^{3} \cong {\rm SU}(2)$. It is useful to note that Eq.~(\ref{scattering-LZ-U}) can be represented in the form of Eq.~(\ref{scattering-LZ-Sp}) by introducing $u (\bm{n}) = n_{+}$, so that $u(\bm{n})$ denote the maps $u : S^{1} \to {\rm U}(1)$ and $u : S^{3} \to {\rm SU}(2)$ for the unitary and symplectic cases, respectively, and in both cases the degree of the relevant map is ${\rm deg} \, (u) = 1$.

Since the expressions in Eqs.~(\ref{scattering-LZ-U}) and (\ref{scattering-LZ-Sp}) are represented in a diabatic basis set, the diagonal and off-diagonal elements of the $2 \times 2$ and block $2 \times 2$ matrices describe the non-adiabatic and adiabatic processes, respectively, so that that the wavepacket components that changes the adiabatic surface does not show any dependence on the direction $\bm{n}$ of the impact parameter, whereas the counterpart that stays on it shows a topologically nontrivial dependence on $\bm{n}$, which will be discussed in some detail in section~\ref{sec:toplology-scatrd}.

\section{Topological Properties of a Scattered Wavepacket}
\label{sec:toplology-scatrd}

In this section we obtain analytical expressions for the wavepacket, right after passing the CS, with focus on its polarization structure, and study the topological properties of the latter. We start with deriving an explicit expression for the scattered wavepacket,which can be readily obtained by substituting Eq.~(\ref{scattering-LZ-U}) or Eq.~(\ref{scattering-LZ-Sp}) into Eq.~(\ref{solution-ballistic}), as explained in section~\ref{sec:conical-semicl-half-int}.

Indeed, let $\bar{\Psi}_{1} (\bm{x}; z) = \bar{\Psi}_{1} (\bm{n}, x; z)$ be the incident wavepacket at the initial time $t_{0}$; the coordinates are relative to the wavepacket position that by definition lies on the ballistic trajectory that goes exactly through the conical point, which means that the position is completely defined by $z_{1} < 0$, so that the actual position of a configuration in the wavepacket is $(\bm{x}; z_{1} + z)$. Note that if a wavepacket has a well-defined center, e.g., in the Gaussian case, the position is generically shifted with respect to the center by the impact parameter of the ballistic trajectory, associated with the center. Let $z_{2} > 0$ be the position of the scattered wavepacket, at time $t$, with the obvious relation $z_{2} = z_{1} + v (t - t_{0})$, and let $\bar{\Psi}_{2} (\bm{x}; z) = \bar{\Psi}_{2} (\bm{n}, x; z)$ be the scattered wavepacket, with the coordinates naturally defined relative to the new position.

Being focused on a more interesting case of the wavepacket staying on an adiabatic surface we obtain, e.g., for the upper adiabatic surface
\begin{eqnarray}
\label{WP-scatrd-ad-upper} \bar{\Psi}_{2} (\bm{n}, x; z) &=& e^{iS_{\rm B} / \hbar} \sqrt{1 - P_{\rm d} (\sqrt{2}x/r_{\rm s})} e^{-i \Phi_{a}} \nonumber \\ &\times& u(\bm{n}) \bar{\Psi}_{1} (\bm{n}, x; z),
\end{eqnarray}
with
\begin{eqnarray}
\label{WP-scatrd-ad-upper-2} \Phi_{\rm a} &=& \Phi (\sqrt{2}(z_{1} + z)/r_{\rm s}; \sqrt{2}x/r_{\rm s}) \nonumber \\ &+& \Phi (\sqrt{2}(z_{2} + z)/r_{\rm s}; \sqrt{2}x/r_{\rm s}) - \delta (\sqrt{2}x/r_{\rm s}),
\end{eqnarray}
so that evaluating the r.h.s. of Eq.~(\ref{WP-scatrd-ad-upper}) we arrive at the following explicit expression
\begin{eqnarray}
\label{WP-scatrd-ad-upper-3} && \bar{\Psi}_{2} (\bm{n}, x; z) = \nonumber \\ && \times e^{iS_{\rm cl} / \hbar} e^{-i(z_{1} + z_{2})z/r_{\rm s}^{2} - i(x^{2}/(2r_{\rm s}^{2})) \ln (2|z_{1}z_{2}|/r_{\rm s}^{2})} \nonumber \\ && \times \sqrt{1 - \exp (- \pi x^{2}/r_{\rm s}^{2})} e^{i \pi/4 - i \, {\rm arg} \, \Gamma (-i x^{2}/(2r_{\rm s}^{2})) - iz^{2}/r_{\rm s}^{2}} \nonumber \\ && \times u(\bm{n}) \bar{\Psi}_{1} (\bm{n}, x; z),
\end{eqnarray}
with the classical action
\begin{eqnarray}
\label{WP-scatrd-ad-upper-4} S_{\rm cl} = \frac{mv^{2} (t - t_{0})}{2} - \frac{fz_{1}^{2} + fz_{2}^{2}}{2}v.
\end{eqnarray}

The final expression for the wavepacket scattering [Eqs.~(\ref{WP-scatrd-ad-upper-3}) and (\ref{WP-scatrd-ad-upper-4})] can be interpreted in the following way. The scalar and matrix factors in the second and third lines of the r.h.s. of Eq.~(\ref{WP-scatrd-ad-upper-3}) are independent of the initial $z_{1}$ and final $z_{2}$ positions and describe strong non-adiabatic effects, associated with the wavepacket passing through the conical seam. The action $S_{\rm cl}$ is easily identified as the action, associated with a classical particle of mass $m$ ballistic propagation exactly through the conical point in the potential $V(z) = f|z|$ of the upper adiabatic surface, taken in the diabatic approximation. The remaining factor in the first line of the r.h.s. provides a $z_{1}$- and $z_{2}$-dependent correction to the wavepacket momentum, and a Gaussian correction to its shape, represented by the first and second terms in the exponent, respectively. They are responsible for the semiclassical adiabatic dynamics of the wavepacket in the matching region $r_{\rm s} \ll r \ll r_{\rm B}$, where the ballistic approximation also holds. This factor plays an important role in connecting the wavepacket dynamics in the adiabatic and ballistic region, ensuring the independence of the final result on a particular choice of the intermediate points $z_{1}$ and $z_{2}$, as long as both belong to the matching region.

We reiterate that, as observed earlier, Eq.~(\ref{WP-scatrd-ad-upper-3}) describes both the unitary and symplectic cases by interpreting $u$ as $u : S^{1} \to {\rm U} (1)$ and $u : S^{3} \to {\rm SU} (2)$. We further note that the orthogonal case also fits the aforementioned expression by setting $u : S^{0} \to \mathbb{Z}_{2}$ to an identity map, making use of $S^{0} = \{-1, 1\} = \mathbb{Z}_{2}$.

\begin{figure}[!ht]
	\includegraphics[width=0.45\textwidth]{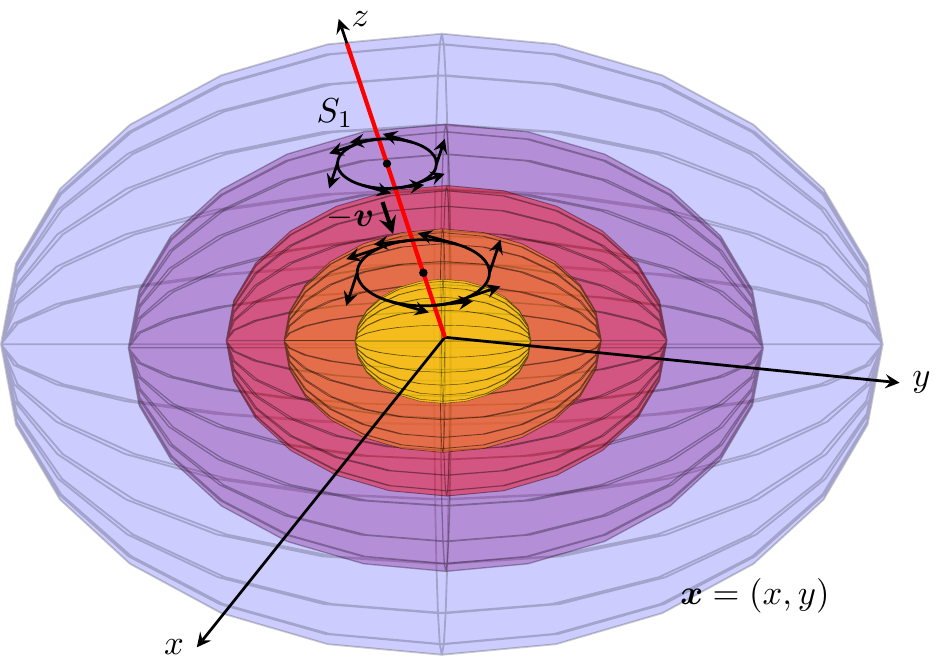}
	\caption{Nontrivial topological structure of the scattered wavepacket. The red line represents the conical trajectory that, in the proper frame, associated with the wavepacket, pinches the wavepacket along $z$\protect\nobreakdash-axis, where the wavefunction turns to zero (nodal line). The polarization vector performs a complete $2\pi$ rotation along any circle that surrounds the conical trajectory (two are shown).}
	\label{fig:spin-conical-trajectory}
\end{figure}

We are now in a position to identify the topological properties of the scattered wavepacket that are completely determined by the matrix factor in the last line of Eq.~(\ref{WP-scatrd-ad-upper-3}). We start with the simpler unitary case in its minimal dimension $d = 3$. In the frame, moving together with the wavepacket, hereafter referred to as the proper frame, the conical point moves with a constant velocity $-\bm{v}$, pinching the wavepacket along a segment of a straight line, hereafter referred to as the conical trajectory, as shown in Fig. \ref{fig:spin-conical-trajectory}. According to the earlier agreement the wavepacket position should be chosen as a point that belongs to the conical trajectory. The $z$-axis in Fig. \ref{fig:spin-conical-trajectory} is not orthogonal to the $\bm{x} = (x, y)$ plane, since, as explained in section~\ref{sec:conical-semicl-half-int}, we use a coordinate system that diagonalizes the matrix of the force constants at the conical point, rather than the mass matrix $m_{ij}$, with the second one usually referred to as the reduced coordinate system. By the same reason the lines of constant values of $z$ and $x = |\bm{x}|$ appear to be ellipses, rather than circles; however they are still circles topologically and therefore will be denoted $S^{1}$. Recalling our definition of polarization, given at the end of section~\ref{sec:BO-semicl-half-int} for the symplectic case, adopting it to the unitary case, and applying it to $\bar{\Psi}$, rather than $\Psi$, with the two related via Eq.~(\ref{Psi-ballistic}), we have $\bar{\Psi} (\bm{x}, z) = \zeta (\bm{x}, z) |\bar{\Psi} (\bm{x}, z)|$, and further observe from Eq.~(\ref{WP-scatrd-ad-upper-3}) that, if the polarization of the incident wavepacket is $(\bm{x}, z)$-independent, than the phase of $\zeta$ acquires $2\pi$ upon performing a full rotation over the circle $S^{1}$, reflecting the fact that the degree of the map $\zeta : S^{1} \to {\rm U}(1)$ is ${\rm deg} \, \zeta = 1$. The fact that the degree of a map is a topological (strictly speaking, homotopy) invariant, makes it robust. In particular, we will still have ${\rm deg} \, \zeta = 1$ for any, generically curved path that winds along the trajectory of the conical point once. Secondly, the topologically nontrivial structure of the scattered wavepacket will still be in place if the initial polarization is not necessarily homogeneous, but also in the case when its phase is well-defined, which happens, e.g., in the case when the wavefunction does not have zeros within its support. This is true, e.g., for a very relevant example of a Gaussian wavepacket, and not true for the scattered counterpart that has zeros on the conical point trajectory. Third, if one finds even a single circle with the nontrivial associated ${\rm deg} \, \zeta = 1$, this immediately implies that the wavefunction will turn to zero on some line within the wavepacket, which follows from the argument that the wavefunction should turn to zero at at least one point on any disc, spanned onto the circle. The latter follows from a standard topological argument and is intuitively obvious.

\begin{figure}[!h]
	\includegraphics{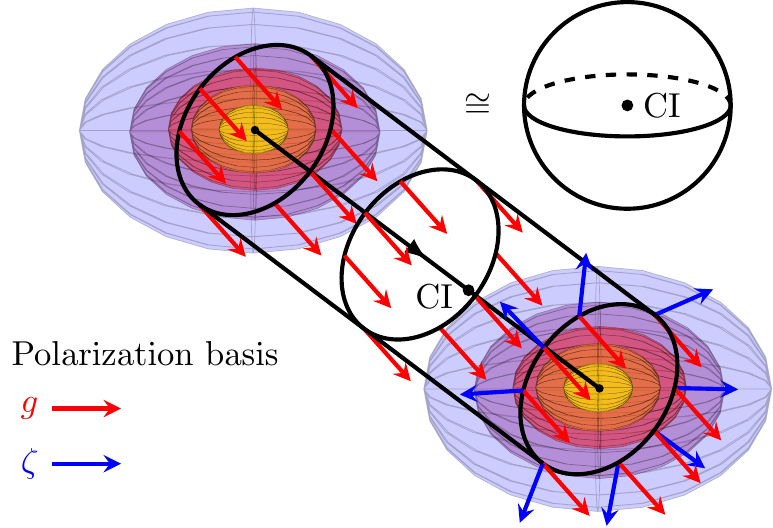}
	\caption{Illustration of the argument that connects the topologically non-trivial structure of the scattered wavepacket to the Chern class, associated with a CS, represented by a single point. Red vectors show the wavepacket polarization in the adiabatic region, using an appropriate adiabatic basis set that does not have singularities inside the initial wavepacket. In the adiabatic region polarization is preserved by ballistic evolution. Blue vectors show the final polarization using an alternative and appropriate basis set that is regular within the scattered wavepacket. By the topological argument polarization in the appropriate basis set shows nontrivial topological structure.}
	\label{fig:spin-topology}
\end{figure}

The topologically non-trivial structure of the scattered wavepacket, namely, ${\rm deg} \, u =1$, for the map $u$, associated with a circle, that winds around the conical trajectory, is directly related to the non-trivial value $c_{1} = 1$ of the first Chern class, which represents the relevant topological invariant associated with conical seams in the unitary case. An argument that demonstrates the aforementioned relation is illustrated in Fig. \ref{fig:spin-topology}. It is based on considering a circle that lies inside the wavepacket in its adiabatic region, and winds around the conical trajectory, e.g., by fixing the value of $z$, say to $z = 0$. In the laboratory frame, upon ballistic motion of the wavepacket, this circle will span a cylinder, as shown in Fig. \ref{fig:spin-topology}. Spanning $2$-dimensional discs $D^{2}$ on the initial and final circles we obtain a surface, topologically equivalent to $S^{2}$, that winds around the conical point, and therefore, the upper adiabatic level, associated with the surface, has Chern class $c_{1}=1$. Fixing the phase of the adiabatic state on the initial disc according to the actual wavefunction, we can then extend it to the cylinder by applying adiabatic propagation, resulting in a well defined basis, defined on the surface, except for the final disc. As for the final disc, it is natural to fix the phase to be position independent. By the arguments, presented in section~\ref{sec:conical-topol-inv}, the latter basis set, being restricted to the circle is connected to its counterpart, restricted from the cylinder, i.e., obtained from solving the dynamical problem, via a map $g : S^{1} \to {\rm U}(1)$ with degree ${\rm deg} \, g = c_{1}$, so that the topological structure of the final wavepacket, namely ${\rm deg} \, \zeta = {\rm deg} \, g = 1$ for its polarization $\zeta$, is determined by the value $c_{1}=1$ of the topological invariant, associated with the conical seam.

The more relevant symplectic case is analyzed in an absolutely similar way. In fact all arguments, presented above for the unitary case, stay conceptually the same, with just a couple of differences in details. Namely, the minimal space dimension should be changed from $d=3$, to $d=5$, the circles $S^{1}$ that surrounds the conical trajectory and the discs $D^{2}$, spanned on them, are replaced by the $3$-dimensional spheres $S^{3}$ and $4$-dimensional discs $D^{4}$, respectively. Also the map $g : S^{1} \to {\rm U}(1)$ and the related first Chern class $c_{1}$ are replaced by $g : S^{3} \to {\rm SU}(2)$ and second Chern class $c_{2}$, as follows from material, presented in section~\ref{sec:conical-topol-inv}. The aforementioned strong similarity of the two cases (which borders with identity, at least in the conceptual sense), together with dealing with much more intuitive $3$-dimensional case, compared to $5$-dimensional counterpart, was the actual reason why we chose to focus on the unitary case in our presentation.

We are now in a position to briefly discuss the topology of the ballistic case for arbitrary dimension $d \ge 3$ and $d \ge 5$, for the unitary and symplectic cases, respectively, now focusing on the symplectic situation, with the unitary being interpreted by analogy. The local coordinate system is chosen by slightly modifying the approach, presented in section~\ref{sec:conical-semicl-half-int} for the minimal dimension $d=5$ case. We chose the $z$ axis along the velocity direction and achieve $\hat{h}(\bm{e_{5}}) = f \gamma_{5}$ in exactly the same way. We further choose some orthonormal, with respect to the mass-weighted scalar product, basis set $(\bm{e}_{k} \, | \, 6 \le k \le d)$ along the conical seam. We further impose the conditions $\hat{h} (\bm{e}_{j}) = f \gamma_{j}$, for $j = 1, \ldots, 4$, which, together with the requirement of orthogonality to the conical seam, completely identify $(\bm{e}_{j} \, | \, 1 \le j \le 4)$. It is natural to denote the corresponding coordinate components $\bm{x} = (x_{1}, x_{2}, x_{3}, x_{4})$ The conical trajectory becomes a $(d-4)$-dimensional plane that pinches the wavepacket along a $(d-4)$-dimensional disc $D^{d-4}$, where the wavepacket position should be chosen, and where the wavefunction of the scattered wavepacket turns to zero. This disc can be winded by $3$-dimensional spheres, e.g., restricting them to the $\bm{x}$ spaces, so that on each of these spheres the polarization is topologically non-trivial, same as in the minimal dimension case. On a more general note, in the semiclassical/ballistic approximation, with the aforementioned coordinate choice the wavepacket evolves only along the $5$ essential coordinates $(\bm{x}, z)$ with nothing happening to its dependence on the rest of coordinates, chosen along the conical seam.

The topological nature of the scattered wavepacket structure is an important observation, due to robustness of topological features with respect to continuous parameter changes, which implies that when the parameter $g_{\rm s}$ becomes larger, so that the ballistic approximation does not hold quantitatively, the main features, i.e., the wavefunction turning to zero on some $(d-4)$-dimensional disc, generically curved, and the topological feature of the polarization around it, will preserve, at least in the region from small to modest values of $g_{\rm s}$, providing strong, topologically protected evidence of the wavepacket to have passed through a conical seam.

\section{Conclusion}
\label{sec:discussion}

In this paper we addressed non-adiabatic effects in photoinduced dynamics of molecules with odd number of electrons (radicals), with focus on semiclassical treatment. Similar to~\cite{Piryatinski2005}, where spin has been not considered at all, we built a semiclassical theory that accounts for non-adiabatic transitions, which is asymptotically exact in the $\hbar \to 0$ limit. Similar to the simpler integer spin case, in the proper semiclassical limit non-adiabatic transitions occur only in the neighborhood of the conical seam, whose transverse size is given by the scattering length $r_{\rm s}$. In our earlier work non-adiabatic transitions have been accounted for via modification of the Van Vleck semiclassical propagator, in the region where a classical trajectory passes by the conical seam. Here we developed an equivalent, still more intuitive approach, formulated using wavepacket dynamics in the following way. While far away from the conical seam, a wavepacket moves adiabatically and semiclassically, according to Van Vleck picture, in particular preserving a Gaussian shape. The conical seam is passed ballistically, with the wavepacket experiencing completely local changes, according to a multistate (in the half-integer spin case $4$-state) Landau-Zener evolution.

There are still some important differences, implied by time-reversal symmetry, in particular Kramers permanent degeneracy of the electronic levels/potential energy surfaces. To identify the dynamical consequences of the aforementioned permanent degeneracy and interpret them in a clear and intuitive way, in section~\ref{sec:kramers-symmetry} we have formulated time-reversal symmetry using proper terms, reformulating the results of Mead \cite{Mead1979,Mead1987} and Matsika-Yarkony \cite{doi:10.1063/1.1378324,doi:10.1063/1.1391444,doi:10.1063/1.1427914,doi:10.1021/jp020396w} in a form ready for dynamical implementation. In particular, representing the electronic Hamiltonians in the vicinity of unavoidable crossings as a linear combination of gamma-matrices, represented by $\sigma_{z}$ and $\sigma_{x}$, all three Pauli matrices, and four Dirac gamma matrices together with $\gamma_{5}$, in the orthogonal (integer spin), unitary (no time-reversal symmetry), and symplectic (half-integer spin) cases, respectively, allowed later (in section~\ref{sec:conical-semicl-half-int}) the ballistic propagation to be treated within the same framework, in particular express the results in terms of a ''standard'' $2 \times 2$ Landau-Zener problem, by making use of the gamma-matrix algebra.

We have identified the symplectic group ${\rm Sp} (1)$ as the one being responsible for Kramers degeneracy, and pointed to its isomorphism to special unitary group ${\rm SU} (2)$, the latter being more common in the chemical physics community. In section~\ref{sec:BO-semicl-half-int} we have extended the Born-Oppenheimer approximation to the permanent degeneracy case, and have demonstrated that, in the semiclassical limit, the wavefunction polarization that defines the value of the function in the double-degenerate electronic space, and is represented by a unit length $2$-component complex vector, evolves completely geometrically, according to parallel transport, the latter effect leading to a non-abelian Berry phase, represented by an ${\rm SU} (2)$ rotation, if one moves over a close loop trajectory.

We further demonstrated, using the ballistic approximation, that once completely passed through a conical seam, the wavepacket component that stays on the same adiabatic surface adopts a topologically non-trivial structure: the wavefunction turns to zero on a $(d-4)$-dimensional surface, represented by the points in the wavepacket that went exactly through the $(d-5)$-dimensional CS, and that in the transverse directions the polarization shows a topologically nontrivial structure. We have demonstrated that the latter is directly related to the topological invariant of CSs in the symplectic case, namely the second Chern class $c_{2}$, whereas in the orthogonal and unitary cases the corresponding invariant, responsible for the wavepacket structure, is represented by the first Stiefel-Whitney $w_{1}$ and first Chern class $c_{1}$, respectively. Such identification is an important observation since, due to robust character of topology, the structure, described above, will not disappear when the semiclassical/ballistic approximation is no longer valid, just the aforementioned $(d-4)$-dimensional node surface will get curved, so that the topologically non-trivial polarization structure can be viewed as a strong experimental evidence of the wavepacket to have passed through a CS, for the measurements, sensitive to the wavefunction polarization \cite{doi:10.1021/jz401982a}, e.g., in spin-sensitive fragment angular distributions upon photo-dissociation of half-integer spin radicals. We reiterate that, in the integer spin case, state-of-art numerically exact propagation of nuclear wavepackets with non-adiabatic effects accounted for explicitly, combined with the Landau-Zener spirit analysis showed the sensitivity of photo-dissociation data, available from experiments to the specific details of the wavepacket shape, characteristic to passing through a conical seam, as well as an excellent agreement between the Landau-Zener analysis and numerically exact results \cite{doi:10.1063/1.5019735,doi:10.1063/1.5019738}. Furthermore, an apparently more complicated case of triple-state crossing have been studied for both half-integer and integer spin systems \cite{Matsika2003a,Matsika2003,doi:10.1063/1.1580092,doi:10.1021/ja043093j}. It is worth mentioning that, the CS in a triple crossing integer spin system has codimension $5$ which equals to the CS in a double crossing half-integer spin system, despite the different local structures of their Hamiltonian in the vicinity of the CSs.

Obviously the ballistic approximation allows not only the shapes of the scattered wavepackets to be determined, but also the evolution of the complete wavefunction during whole the scattering process to be followed. This can be easily achieved by replacing the limiting values of the matrix elements in Eq.~(\ref{evolution-LZ}), given by Eq.~(\ref{scattering-LZ}), with the actual values, expressed in terms of the parabolic cylinder functions, as presented in~\cite{Piryatinski2005}. These should provide a clear semiclassical interpretation of the recently proposed time-resolved X-ray experiments \cite{kowalewski2017monitoring,doi:10.1021/acs.chemrev.7b00081}, capable of providing detailed dynamical information on the nuclear wavepacket passage through a CS. It is worth noting that in the proper adiabatic (i.e., diabatic) basis set, associated with the reference trajectory, the wavepackets that stay and change the diabatic surface, will have a topologically plain and topologically non-trivial polarization structure, respectively.

Finally, it would be of interest to explore a possibility of combining the presented semiclassical dynamical view of scattering at conical intersections with widely used surface hopping algorithms, especially the ones that properly account for the quantum phase effects, see, e.g., \cite{Gorshkov2013,White2014}, to improve their performance in the situation when conical intersections are involved.

\acknowledgments
This material is based upon work supported by the National Science Foundation under Grant No. CHE-1111350. We acknowledge support of Directed Research and Development Funds, Center for Integrated Nanotechnology and Center for Nonlinear Studies at Los Alamos National Laboratory (LANL). LANL is operated by Los Alamos National Security, LLC, for the National Nuclear Security Administration of the U.S. Department of Energy under contract DE-AC52-06NA25396. We thank Michael J. Catanzaro, John R. Klein, Nikolai A. Sinitsyn, and Sergei Tretiak for useful discussions. We also wish to thank the anonymous reviewer for an outstanding job, constructive criticism, and important comments that allowed us to substantially improve the manuscript.

\appendix

\section{Scalar Products, Symplectic Forms, And Symplectic Groups}
\label{sec:Sp-groups}

In this appendix we present certain notation, definitions, and properties of {\it symplectic} groups, together with some derivations.
One of the reasons we wrote this appendix is that there is an ambiguity in the notation used by several various sources.

The symplectic group ${\rm Sp}(2m;\mathbb{F})$, with $\mathbb{F}$ being a field, is the group of invertible linear operators acting in the $2m$-dimensional vector space $V \cong \mathbb{F}^{2m}$, equipped with a symplectic form $\omega$, preserved by the aforementioned linear operators. Usually the cases $\mathbb{F}= \mathbb{R}$ or $\mathbb{F}= \mathbb{C}$ are considered. We will focus on the case ${\rm Sp}(2m;\mathbb{C})$ that is relevant for our applications. A natural question arises: why and in what way are symplectic groups closely related to time-reversal symmetry in quantum systems? The answer can be formulated as follows. By definition, a symplectic form is just a non-degenerate bilinear form in a (complex) vector space $V$, which is skew-symmetric, i.e., it satisfies the property $\omega(u\otimes v)= -\omega(v\otimes u)$, for $u, v \in V$.


We further observe that a Hermitian scalar product, which is always a part of a game for a quantum system, establishes a one-to-one correspondence between antilinear maps and bilinear forms (not necessarily skew-symmetric) that is uniquely determined by the condition
\begin{eqnarray}
\label{j-to-omega} \omega(u\otimes v)= (u, j(v)) \;\;\; \forall u,v\in V.
\end{eqnarray}
We will say that $\omega$ is compatible with the scalar product if the corresponding antilinear map $j$ preserves the scalar product in the sense of Eq.~(\ref{preserve-scalar}). In this case we have
\begin{eqnarray}
\label{omega-skew} && \omega(u\otimes v)= (u, j(v))= (j(u), j^{2}(v))^{*} \nonumber \\ && = (j^{2}(v), j(u))= \omega(j^{2}(v)\otimes u),
\end{eqnarray}
which implies that in our compatible case the skew-symmetry of $\omega$ is equivalent to $j^{2}= -1$. Therefore, there is a one-to-one correspondence between the symplectic forms compatible with the scalar product and $j^{2}= -1$ real structures that preserve the scalar product. Stated differently, and in more physics terms, we are considering a situation when time-reversal symmetry respects the scalar product, the latter being the most important structure in quantum mechanics.

In view of the above we can define a symplectic group ${\rm Sp}(m)$, also often referred to as a {\it compact symplectic} group (since it is in fact compact), in the following way. Let $V$ be a complex vector space of even dimension $2m$, equipped with a Hermitian scalar product and a symplectic form, compatible with the scalar product (or equivalently a $j^{2}= -1$ real structure that preserves the scalar product). The group ${\rm Sp}(m)$ then consists of all linear operators $A$ acting in $V$ that are unitary and preserve the symplectic form (or, equivalently, commute with the corresponding real structure):
\begin{eqnarray}
\label{define-Sp-m} && (A(u), A(v))= (u, v), \;\;\; \omega(A(u)\otimes A(v))= \omega(u\otimes v), \nonumber \\ && \;\;\;\; jA= Aj.
\end{eqnarray}
Stated in more physics terms operators that belong to ${\rm Sp}(m)$ represent unitary operators (i.e., a quantum version of variable changes) that respect time-reversal symmetry.

At this point we would like to note that sometimes ${\rm Sp}(m)$ are denoted ${\rm USp}(2m)$ to emphasize that it is isomorphic ${\rm Sp}(m) \cong {\rm U}(2m) \cap {\rm Sp}(2m;\mathbb{C})$ of unitary symplectic matrices with complex entries. This definition is somewhat sloppy due to the following reason. The notion of unitarity can exist only if a Hermitian scalar product is defined. The symplectic group can be properly defined only if the symplectic form $\omega$ is compatible with the scalar product in the sense, explained above. The compatibility condition is also often dropped out of the definition of the ${\rm Sp}(m)$ groups, since they are usually defined in terms of matrices, using an orthonormal basis set in which the symplectic form has a standard (canonical) form
\begin{eqnarray}
\label{omega-canonical} \omega= \left(\begin{array}{cc} 0 & I \\ -I & 0 \end{array}\right),
\end{eqnarray}
with $I$ being the unit $m\times m$ matrix, and it can be straightforwardly verified that the canonical symplectic form [Eq.~(\ref{omega-canonical})] is compatible with the canonical scalar product (associated with an orthonormal basis set).

There is another standard, and also very convenient model for the ${\rm Sp}(m)$ group, referred to as the unitary quaternionic group $U(m;\mathbb{H})$ that consists of all invertible $m\times m$ matrices with quaternionic entries that preserve the standard Hermitian scalar product
\begin{eqnarray}
\label{scalar-quaternion} \langle \bm{u}, \bm{v}\rangle= \sum_{a=1}^{m} u_{a}v_{a}^{*} \in \mathbb{H}.
\end{eqnarray}
The isomorphism ${\rm Sp}(m) \cong U(m;\mathbb{H})$ can be established by using {\it real orthonormal} basis sets, i.e., orthonormal basis sets of a form $(e_{1},\ldots, e_{m}, j(e_{1}),\ldots, j(e_{m}))$. Such basis sets can be actually built by applying an obvious extension of the Gram-Schmidt orthogonalization procedure that on each step builds a new pair $(e_{a},j(e_{a}))$ of the basis set elements, orthogonal to the previously chosen ones. One then can choose $(e_{1},\ldots, e_{m})$ as the basis set, forming the $m$-dimensional quaternionic space $\mathbb{H}^{m}$, to represent the linear operators acting in $V$ that commute with the real structure $j$ using $m\times m$ quaternionic matrices and show directly that an operator $A$ preserves a Hermitian scalar product in the $2m$-dimensional complex vector space $V$ if and only if the corresponding $m\times m$ quaternionic matrix preserves the quaternionic scalar product, given by Eq.~(\ref{scalar-quaternion}).

A quaternionic scalar product in a complex vector space $V$ equipped with a Hermitian scalar product and a real structure that preserves the latter can be introduced in an invariant way
\begin{eqnarray}
\label{scalar-quaternion-2} \langle u, v \rangle= (u,v)+ (u, J(v))j, \;\;\; \langle u, v \rangle \in \mathbb{H};
\end{eqnarray}
here for the sake of clearness of the derivations, presented below we do not overload the notation for $j$, by still denoting with $j$ the element of the quaternion algebra $j\in \mathbb{H}$, while using $J$ for the real structure anti-linear map $J: V \to V$. The introduced scalar product has the following important properties. First, and though obvious, still very important: the quaternionic scalar product $\langle u, v \rangle$ provides two Hermitian scalar products $(u,v)$ and $(u, J(v))$. Second, for an invertible operator $A$ the property of preserving the quaternionic scalar product is equivalent to preserving the Hermitian scalar product and the real structure, the latter meaning $[J,A]= 0$. This can be demonstrated as follows. The preservation of the quaternionic scalar product means
\begin{eqnarray}
\label{slalar-quaternion-preserve} && (A(u), A(v))= (u,v), \nonumber \\ && (A(u), JA(v))= (u, J(v)), \;\;\; \forall u,v\in V.
\end{eqnarray}
The first relation means preservation of the Hermitian scalar product, whereas applying the first relation to the r.h.s. of the second one we obtain
\begin{eqnarray}
\label{slalar-quaternion-preserve-2} (A(u), JA(v))= ((A(u), AJ(v)), \;\;\; \forall u,v\in V,
\end{eqnarray}
which is equivalent to $AJ= JA$. Finally the following bilinear properties are in place
\begin{eqnarray}
\label{scalar-quaternion-3} \langle \lambda u,v \rangle= \lambda \langle u,v \rangle, \;\;\; \langle u,\lambda v \rangle= \langle u,v \rangle \lambda^{*},
\end{eqnarray}
for $u,v\in V$ and $\lambda\in \mathbb{H}$. To verify the properties, presented in Eq.~(\ref{scalar-quaternion-3}), it is enough to verify them for $\lambda \in \mathbb{C}\subset \mathbb{H}$ and for $\lambda= j$. For $\lambda \in \mathbb{C}$ we have
\begin{eqnarray}
\label{scalar-quaternion-lin-1} \langle \lambda u,v \rangle &=& (\lambda u,v)+ (\lambda u, J(v))j \nonumber \\ &=& \lambda(u,v)+ \lambda(u, J(v))j= \lambda\langle u,v \rangle, \nonumber \\ \langle u,\lambda v \rangle &=& (u,\lambda v)+ (u, \lambda^{*}J(v))j \nonumber \\ &=& \lambda^{*}(u,v)+ \lambda(u, J(v))j \nonumber \\ &=& (u,v)\lambda^{*}+ (u, J(v))j\lambda^{*} \nonumber \\ &=& \langle u,v \rangle\lambda^{*},
\end{eqnarray}
whereas for $\lambda= j$
\begin{eqnarray}
\label{scalar-quaternion-lin-2} 
\langle ju,v \rangle
&=& \langle J(u),v \rangle =  (J(u),v)+ (J(u), J(v))j \nonumber \\ 
&=& -(u, J(v))^{*}+ (u,v)^{*}j \\ 
&=& j^{2}(u, J(v))^{*}+ (u,v)^{*}j \nonumber \\ 
&=& j(u, J(v))j+ j(u,v)= j\langle u,v \rangle, \nonumber \\ 
\langle u,jv \rangle 
&=& \langle u,J(v) \rangle =  (u, J(v)) + (u, J^{2}(v))j \nonumber \\ 
&=& -(u, v)j+ (u, J(v)) \\ &=& -(u, v)j- (u, J(v))j^{2} \nonumber \\ 
&=& -\langle u,v \rangle j= \langle u,v \rangle j^{*}. \nonumber
\end{eqnarray}
The bilinear properties [Eq.~(\ref{scalar-quaternion-3})] imply that if the vectors $u,v\in V$ are decomposed using an orthonormal, with respect to the quaternionic scalar product, quaternionic basis set, and quaternionic coefficients $u_{a},v_{a}\in \mathbb{H}$
\begin{eqnarray}
\label{u-v-decompose} u= \sum_{a= 1}^{m}u_{a}e_{a}, \;\;\; v= \sum_{a= 1}^{m}v_{a}e_{a}, \;\;\; \langle e_{a}, e_{b} \rangle= \delta_{ab},
\end{eqnarray}
so that the quaternionic scalar product has a form of Eq.~(\ref{scalar-quaternion}).

We further describe the notion of a linear operator $H$ being quaternionically Hermitian, which naturally reads
\begin{eqnarray}
\label{Hermition-quaternion} \langle H(u), v \rangle &=& \langle u, H(v) \rangle, \;\;\; \forall u,v\in V,
\end{eqnarray}
or explicitly
\begin{eqnarray}
\label{Hermition-quaternion-a} && (H(u), v) + (H(u), J(v))j \nonumber \\ && \;\;\; = (u, H(v)) + (u, J(H(v)))j,
\end{eqnarray}
or recasting further in components
\begin{eqnarray}
\label{Hermition-quaternion-b} && (H(u), v) = (u, H(v)), \nonumber \\ && \;\;\; (H(u), J(v))= (u, J(H(v))).
\end{eqnarray}
The first equality in Eq.~(\ref{Hermition-quaternion-b}) simply means that $H$ is Hermitian; applying it to the second one we arrive at
\begin{eqnarray}
\label{Hermition-quaternion-2} (u, HJ(v))= (u, J(H(v))), \;\;\; \forall u,v\in V,
\end{eqnarray}
which implies that $[J, H]= 0$. Summarizing, quaternionically Hermitian operators are exactly Hermitian operators that commute with $J$, i.e., preserve the real structure.

Since quaternions do not commute we need to describe carefully how to represent linear operators in the matrix form. Consider a linear operator $H$ that preserves the real structure, hereafter referred to as a quaternionic operator that has the property $H(\lambda u)= \lambda H(u)$ for $\lambda\in \mathbb{H}$. For a quaternionic orthonormal basis $(e_{1}, \ldots, e_{m})$ we can define the matrix elements with the condition
\begin{eqnarray}
\label{matrix-elem-quaternion} H(e_{b})= \sum_{b= 1}^{m}H_{ab}e_{a}, \;\;\; H_{ab}\in \mathbb{H},
\end{eqnarray}
so that
\begin{eqnarray}
\label{matrix-elem-quaternion-2} && H(u)= H\left(\sum_{b}u_{b}e_{b}\right)= \sum_{ab}u_{b}H_{ab}e_{a}, \nonumber \\ && H(u)_{a}= \sum_{b}u_{b}H_{ab}, \;\;\; u_{a}, u_{b} \in \mathbb{H},
\end{eqnarray}
which means that we can use standard matrix representation with the order of multiplication of the matrix elements with the vector components, prescribed by Eq.~(\ref{matrix-elem-quaternion-2}).

We conclude the discussion of the quaternionic scalar product by recalling a statement that orthonormal in the quaternionic sense [Eq.~(\ref{u-v-decompose})] basis sets are in one-to-one correspondence with orthonormal real basis sets, introduced earlier: given a quaternionic orthonormal basis set $(e_{1},\ldots, e_{m})$ we can build a real orthonormal basis set $(e_{1},\ldots, e_{m}, J(e_{1}),\ldots, J(e_{m}))$. Note that all basis set elements $e_{1},\ldots, e_{m}, J(e_{1}),\ldots, J(e_{m})\in V$.

It would be instructive to note that the group ${\rm Sp}(m)$ can be viewed as the compact real counterpart of ${\rm Sp}(2m;\mathbb{C})$ in the following sense. The latter group is complex analytical (and naturally, being non-abelian, is non-compact), i.e., as a space it is a complex-analytical manifold. The map $\bar{J}: {\rm Sp}(2m;\mathbb{C}) \to {\rm Sp}(2m;\mathbb{C})$, defined by $\bar{J}(A)= A^{\dagger}$ is a real structure, since it is anti-holomorphic, i.e., transforms holomorphic functions to anti-holomorphic, preserves the group action, and satisfies $\bar{J}^{2}= {\rm id}$. The group ${\rm Sp}(m) \subset {\rm Sp}(2m;\mathbb{C})$ can be considered as the subspace of the real points of ${\rm Sp}(2m;\mathbb{C})$, i.e., the fixed points of $\bar{J}$. An elementary computation, based on identification of the Lie algebras, associated with the above Lie groups shows that ${\rm Sp}(2m;\mathbb{C})$,
has complex dimension $m(2m+ 1)$, whereas ${\rm Sp}(m)$ has real dimension $m(2m+ 1)$, the latter in accordance with ${\rm Sp}(m)$ being the real counterpart of ${\rm Sp}(2m;\mathbb{C})$.

\section{Orthogonal Groups, Spinors, and Gamma-Matrices}
\label{sec:spinors}

In this appendix we present some simple basic facts about spinors, necessary to formulate nice interpretation of the conical intersections. A nice and concise overview of the spinors and gamma-matrices for arbitrary dimension can be found in~\cite{kirillov2012elements}.

The orthogonal groups ${\rm SO}(n)$ with $n \ge 3$ are known to be connected, but not simply connected, the latter meaning that they have a not-contractible cycle. The group of equivalence classes (with respect to homotopy) of one-dimensional closed curves with a given origin in a space $X$ is called its fundamental group, and denoted $\pi_{1}(X)$. As known $\pi_{1}({\rm SO}(n))= \mathbb{Z}_{2}$ for $n \ge 3$. For any topological group $G$ there is a uniquely defined topological group $\tilde{G}$, referred to as the universal cover of $G$, that covers $G$, i.e., $\tilde{G} \to {G}$, with $\pi_{1}(\tilde{G})= 0$ and the fiber, i.e., the inverse image of any point in $G$ with respect to the cover map, being isomorphic to $\pi_{1}(G)$. The universal (double) cover of ${\rm SO}(n)$ is called ${\rm Spin}(n)$, so that we have ${\rm Spin}(n) \to {\rm SO}(n)$. A group ${\rm Spin}(n)$ has a canonical unitary representation, referred to as the {\it spinor representation} and a set of $\gamma$-matrices $(\gamma_{a}|a=1,\ldots, n)$, acting in the space of the spinor representation, that satisfy the Clifford algebra relations
\begin{eqnarray}
\label{gamma-Clifford} \gamma_{a}\gamma_{b}+ \gamma_{b}\gamma_{a}= 2\delta_{ab}.
\end{eqnarray}
Under the action of ${\rm Spin}(n)$ the gamma-matrices transform linearly and preserve the natural (real) scalar product, so that elements of ${\rm Spin}(n)$ are represented by orthogonal operators acting in the $n$-dimensional space $\mathbb{R}^{n}$, spanned on the $\gamma$-matrices, which defines the cover ${\rm Spin}(n) \to {\rm SO}(n)$.

There is a well-known explicit construction for ${\rm Spin}(n)$, the spinor representation and $\gamma$-matrices, which we do not give here, but rather present some basic facts. The group ${\rm Spin}(n)$ for $n= 2m$ and $n= (2m+ 1)$ acts in the same vector space of (complex) dimension $2^{m}$, with the $\gamma$-matrices for $n= (2m+ 1)$ obtained by extending the set of $\gamma$-matrices for $n= 2m$ with the product $\prod_{a= 1}^{2m}\gamma_{a}$. There are the following identifications ${\rm Spin}(3) \cong {\rm SU}(2) \cong {\rm Sp}(1)$, with the well-known cover ${\rm SU}(2) \to {\rm SO}(3)$, and $\gamma$-matrices represented by the Pauli matrices $\bm{\sigma}$. For $n= 4$ we have ${\rm Spin}(4) \cong {\rm SU}(2)\times {\rm SU}(2)$, with the $\gamma$-matrices represented by the Dirac matrices. Finally ${\rm Spin}(5) \cong {\rm Sp}(2)$, with the action of the latter in $V \cong \mathbb{C}^{4}$, equipped with a scalar product and a real structure $j$ that preserves the latter, and the $\gamma$-matrices represented by Hermitian operators that commute with $j$, as described in some detail in appendix~\ref{sec:Sp-groups}.

\section{Differential Forms, Wedge Products, Stokes Theorem, and Chern Classes}
\label{sec:diff-forms-Chern-cl}

In this appendix we present some basic facts and concepts, associated with differential forms, including wedge products and multidimensional Stokes theorem, as well as representation of Chern classes using differential forms, with applications to rationalizing Eq.~(\ref{define-degree-S-3}) and deriving Eq.~(\ref{second-chern-class}), starting with the former. Further details on differential forms, Stokes theorem, vector bundles and connections can be found in \cite{spivak2018calculus}. The original construction of Chern classes, developed by Chern, which uses differential forms, is adopted in this paper and briefly described in this appendix, can be found in \cite{bott2013differential}.

A differential form $A$ (of rank $k$) on space/manifold $X$ is a smooth function on $X$, whose value at any point $x \in X$ is a skew-symmetric poly-linear ($k$-linear) form on the vector space of tangent to $X$ vectors at $x$. Given a system of local coordinates it can be equivalently viewed as an expression
\begin{eqnarray}
\label{diff-form-coord} A = A_{j_{1} j_{2} \ldots j_{k}} (x) dx^{j_{1}} \wedge dx^{j_{2}} \wedge \ldots \wedge dx^{j_{k}},
\end{eqnarray}
where the wedge product involved in Eq.~(\ref{diff-form-coord}) is simply a skew-symmetric product, which just means $dx^{j} \wedge dx^{i} = - dx^{i} \wedge dx^{j}$. We reiterate that throughout this paper we use the Einstein summation convention. A wedge product of forms $A$ and $B$ with ranks $k$ and $l$, respectively, is a differential form $A \wedge B$, naturally defined as
\begin{eqnarray}
\label{diff-form-wedge} A \wedge B &=& A_{i_{1} \ldots i_{k}} (x) B_{j_{1} \ldots j_{l}} (x) dx^{i_{1}} \ldots \wedge dx^{i_{k}} \nonumber \\ &\wedge& dx^{j_{1}} \wedge \ldots \wedge dx^{j_{l}},
\end{eqnarray}
An exterior differential $dA$ of $A$ is a $(k+1)$-rank form, defined also in a very natural way
\begin{eqnarray}
\label{diff-form-d} dA = \frac{\partial}{\partial x^{j}} A_{i_{1} \ldots i_{k}}(x) dx^{j} \wedge dx^{i_{1}} \ldots \wedge dx^{i_{k}},
\end{eqnarray}
with the following easily verifiable properties in place
\begin{eqnarray}
\label{diff-form-prop} A \wedge B &=& (-1)^{kl} B \wedge A, \nonumber \\ d (A \wedge B) &=& dA \wedge B + (-1)^{k} A \wedge dB, \nonumber \\ d^{2} A &=& d (dA) = 0.
\end{eqnarray}
We note that the exterior differential operator can be defined in an invariant, i.e., coordinate-free way, so that in any local coordinate system it reproduces Eq.~(\ref{diff-form-d}). It is done by defining it for functions, i.e., $0$-rank differential forms, as $df = (\partial f / \partial x^{j}) dx^{j}$ and extending it to arbitrary rank by requiring the properties, given by Eq.~(\ref{diff-form-prop}) to be satisfied. Also note, that, due to skew-symmetric character, the maximal rank of a form is given by the space dimension.

If $f : X \to Y$ is a map of manifolds, and $A$ is a form over $Y$, we can introduce a form $f^{*} A$ over $X$, called the pull-back of $A$ along $f$, in a very natural way, as
\begin{eqnarray}
\label{diff-form-pullback} f^{*} A &=& A_{\alpha_{1} \ldots \alpha_{k}}(f(x)) \frac{\partial f^{\alpha_{1}}(x)}{\partial x^{j_{1}}} \ldots \frac{\partial f^{\alpha_{k}}(x)}{\partial x^{j_{k}}} \nonumber \\ &\times& dx^{j_{1}} \wedge \ldots \wedge dx^{j_{k}}.
\end{eqnarray}
Viewing $f$ as a coordinate transformation, Eq.~(\ref{diff-form-pullback}) can be also interpreted as the transformation law for differential forms under coordinate transformations.

One of the reasons why differential forms are so useful is that they are designed to be integrated, and, as opposed to just functions, they do not require an integration measure. Indeed, a maximal rank form can be always represented as $A = A(x) dx^{1} \wedge \ldots \wedge dx^{n}$, with $A(x)$ being a function. On the other hand, as it follows from Eq.~(\ref{diff-form-pullback}), under coordinate change $A(x)$ transforms via the Jacobian $J(x) = \det (\partial f / \partial x)$ of the coordinate transformation. Therefore, one can define an integral of the aforementioned differential form as

\begin{eqnarray}
\label{diff-form-integral} \int_{X} A = \int_{X} A(x) dx^{1} \ldots dx^{n},
\end{eqnarray}
since the r.h.s. of Eq.~(\ref{diff-form-integral}) does not depend on the coordinate choice, as long as the coordinate transformation preserves orientation, i.e., $J(x) > 0$. A careful reader would notice that the given definition works locally; to make it global one can use a standard argument that involves a so-called partition of unity. The bottom line is that the integral of a maximal rank differential form over a compact oriented manifold is well defined.

Most importantly, forms of lower rank can be also integrated over the cycles of the corresponding dimension. Defining a $k$-cycle as a map $f : M \to X$ of a compact oriented $k$-dimensional manifold to our space we define
\begin{eqnarray}
\label{diff-form-integral-cycle} A (f) = \int_{f} A = \int_{M} f^{*} A,
\end{eqnarray}
and also refer to $A(f)$ as the value of $A$ at cycle $f$.

The (multidimensional) Stokes theorem claims that if $M$ is a manifold of dimension $m$ with boundary $\partial M$, obviously of dimension $m-1$, e.g., $(M, \partial M) = (D^{m}, S^{m-1})$, mapped to $X$, via $f : M \to X$ and $A$ is a form of rank $m-1$ on $X$, then
\begin{eqnarray}
\label{Stokes-theorem} \int_{f} dA = \int_{f|_{\partial M}} A,
\end{eqnarray}
where $f|_{\partial M}$ is the restriction of $f$ to the boundary of $M$, and, in particular, for a manifold without boundary, referred to as just a manifold, i.e., $f$ is an $m$-cycle, e.g., $M = S^{m}$, the r.h.s. of Eq.~(\ref{Stokes-theorem}) turns to zero. The standard Stokes theorem is reproduced by setting $(M, \partial M) = (D^{2}, S^{1})$, and $X = \mathbb{R}^{3}$.

A form $A$ is called closed if $dA = 0$, it is called exact if $A = dB$ for some $B$; obviously due to $d^{2} = 0$, any exact form is closed. We say that $A$ is cohomologically equivalent to $B$ if $(A-B)$ is exact. The set of equivalence (cohomology) classes $[A]$ of closed $k$-forms $A$ over $X$ forms a vector space, refereed to as the $k$-th de Rham cohomology of $X$ and is denoted $H^{k} (X)$. Obviously $H^{k} (X) = 0$ for $k > n = {\rm dim} (X)$ For a compact manifold all cohomology spaces are finite-dimensional vector spaces. If $X$ is connected $H^{0} (X) = \mathbb{R}$, and the cohomology classes are represented by constant functions. If $X$ is also orientable $H^{n} (X) = \mathbb{R}$. The correspondence $H^{n} (X) \to \mathbb{R}$ is obtained by integrating an $n$-form $A$ over the manifold $X$, with the result depending on its class $[A]$ only, due to the Stokes theorem (note that any form of maximal rank is closed).

Locally, a gauge field is represented by a $1$-form $A = A_{j} dx^{j}$ that takes values in the space of $n \times n$ matrices, i.e., for any $j$, $A_{j}$ is an $n \times n$ matrix with the entries $A_{j}^{ab}$, the latter could be real or complex numbers. A gauge transformation, associated with a matrix function $g(x)$ has a form
\begin{eqnarray}
\label{gauge-transform} && A \mapsto g^{-1} A g + g^{-1} dg \nonumber \\ && A_{j} \mapsto g^{-1} A_{j} g + g^{-1} \frac{\partial g}{\partial x^{j}}
\end{eqnarray}
Usually the values of $g(x)$ are restricted to special orthogonal, unitary, or special unitary matrices so that $g(x) \in G$, with $G = {\rm SO} (n)$, $G = {\rm U} (n)$, and $G = {\rm SU} (n)$, respectively. In this paper only $G = {\rm U} (1)$ and $G = {\rm SU} (2)$ are involved. When the gauge transformations are restricted to the aforementioned subgroups of the linear groups, the values of $A_{j}$ are restricted to the corresponding Lie algebras (the latter describing infinitesimal group transformations), represented by real antisymmetric, complex anti-hermitian, and complex anti-hermitian with zero trace matrices respectively. The global construction works as follows. If $U, V \subset X$ are any two intersecting neighborhoods with the gauge field represented by forms $A|_{U}$ and $A|_{V}$, then over the intersection $U \cap V$ they are allowed to be related via a gauge transformation,naturally represented by a matrix function $g_{UV} : U \cap V \to G$. Obviously, consistency conditions should be imposed, i.e., for any three intersecting neighborhoods $U, V, W \subset X$ we should have over the intersection $U \cap V \cap W$ the consistency relation $g_{UV} g_{VW} = g_{UW}$ to be satisfied. A set $\{A_{U}\}$ of forms connected over intersections $U \cap V$ via gauge transformations, defined by the  connecting/gluing maps $g_{UV}$, the latter satisfying the aforementioned consistency conditions on all triple intersections $U \cap V \cap W$, will be referred to as a global gauge field. The connecting/glueing data represented by a family $\{g_{U_{\alpha}U_{\beta}} : U_{\alpha} \cap U_{\beta} \to G\}_{\alpha, \beta \in I}$, with $\bigcup_{\alpha \in I} U_{\alpha} = X$, that satisfy the consistency condition, define an object, called a vector fiber bundle, in the following sense. Consider a vector-``function'' on $X$ that is locally a function, with the local functions being glued together via the connection maps. More formally, let $\Psi = \{\Psi_{\alpha} : U_{\alpha} \to {\cal U}\}_{\alpha \in I}$ be a family of functions with the values in a vector space ${\cal U}$, equipped with a Hermitian scalar product, of dimension $n$, referred to as a fiber, and the rank of the bundle, respectively, so that, for any $\alpha, \beta \in I$, we have $\Psi_{\alpha} (x) = g_{\alpha\beta} (x) \Psi_{\beta} (x)$ over $U_{\alpha} \cap U_{\beta}$; here we used abbreviated notation $g_{\alpha\beta}$ for $g_{U_{\alpha} U_{\beta}}$. Then $\Psi$ is called a global section of the vector bundle, associated with the gluing data.

A globally defined gauge field can be interpreted as an object that allows derivatives of global sections to be introduced. Indeed for $\Psi_{\alpha}$ we can define its ``elongated'', or in other words covariant, derivative as a $1$-form $\nabla \Psi_{\alpha}$ with the values in $V$, as
\begin{eqnarray}
\label{long-derivative} && \nabla \Psi_{\alpha} = d \Psi_{\alpha} + A_{\alpha} \Psi_{\alpha} = (\nabla_{j} \Psi_{\alpha}) dx^{j} \nonumber \\ && \nabla_{j} \Psi_{\alpha} = \frac{\partial \Psi_{\alpha}}{\partial x^{j}} + A_{\alpha j} \Psi_{\alpha}.
\end{eqnarray}
It is easy to see that the local definition of covariant derivatives [Eq.~(\ref{long-derivative})] is consistent on all $U_{\alpha} \cap U_{\beta}$ due to the transformation law, determined by gauge transformations [Eq.~(\ref{gauge-transform})], so that the covariant derivative with respect to a gauge field is defined globally. Note that in the way the material is presented here we have a notion of a gauge field and associated with the latter vector bundle. In differential geometry it is usually formulated the other way around, one starts with a notion of a vector bundle and then considers connections in a given vector bundle; with the connection being a term in differential geometry for what a physicist would call a globally defined gauge field.

The curvature $F$ of a gauge field $A$ is defined locally as a matrix-valued $2$-form, i.e., over $U_{\alpha}$, we have
\begin{eqnarray}
\label{define-F} && F_{\alpha} = d A_{\alpha} + \frac{1}{2} [A_{\alpha}, \wedge A_{\alpha}] = F_{\alpha, ij} dx^{i} \wedge dx^{j} \nonumber \\ && F_{\alpha, ij} = \frac{1}{2} \left(\frac{\partial A_{\alpha j}}{\partial x^{i}} - \frac{\partial A_{\alpha i}}{\partial x^{j}} + [A_{i}, A_{j}]\right),
\end{eqnarray}
with the following gluing data on $U_{\alpha} \cap U_{\beta}$
\begin{eqnarray}
\label{gauge-transform-F} F_{\alpha} (x) = g_{\alpha\beta}^{-1} (x) F_{\beta} (x) g_{\alpha\beta} (x),
\end{eqnarray}
so that the curvature can be interpreted as a $2$-form with values in another vector bundle of rank $n^{2}$, and the fiber, represented by the vector space ${\rm End} ({\cal U})$ of linear operators acting in ${\cal U}$, known as the endomorphism bundle, associated with the original counterpart.

Chern classes $c_{k}$, with $k = 1, 2, \ldots$ are invariants of vector bundles over $X$ with $c_{k} \in H^{2k} (X)$, so that each Chern class is a cohomology class. In this paper, to minimize the algebraic topology involved, we will follow the original construction of Chern, i.e., use the de Rham cohomology, defined earlier in this appendix. We start with defining a Chern class $C_{k} (A)$, associated with a gauge field $A$ as a a $2k$-differential form over $X$ that depends on $A$. We further show that that $C_{k} (A)$ is closed, which allows us to introduce the corresponding de Rham cohomology class $[C_{k} (A)] \in H^{2k} (X)$, making $[C_{k} (A)]$ an invariant of a gauge field. We next demonstrate that the cohomology class $[C_{k} (A)]$ does not depend on a particular choice of the gauge field, for given gluing data, or in other words, vector bundle, so that we can define $c_{k} = [C_{k} (A)]$ as invariants of the vector bundle, rather that a gauge field, and refer to them as Chern classes.

The original Chern construction, we have adopted here, is very simple, however, it has a disadvantage: it is hard to see the integer nature of Chern classes, the latter meaning that the integral of a Chern class $c_{k}$ over any $2k$-cycle, defined by Eq.~(\ref{diff-form-integral-cycle}) (and which does not depend on a choice of a particular representative due to Stokes theorem), is an integer. Understanding the aforementioned integer nature requires bringing in the concept of a classifying space, which, for the case of $n$-dimensional complex vector bundles, we are considering here, is denoted ${\rm BU}(n)$. The classifying space is equipped with a preferred bundle over it, called the universal bundle, and any bundle over $X$ may be pulled back from the universal counterpart along some map $f : X \to {\rm BU}(n)$, so that a Chern class $c_{k}$ is pull-backs [in the sense of Eq.~(\ref{diff-form-pullback})] of some integer-valued basis class $\bar{c}_{k} \in H^{2k} ({\rm BU}(n))$, referred to as a Chern class of the universal bundle, or simply a universal Chern class, so that $\bar{c}_{k}$ generate the complete cohomology of the classifying space. The cohomology of the classifying space is well known due to existence of a very simple model ${\rm BU}(n) = {\rm colim}_{N \to} G (n; N+n; \mathbb{C})$, where $G (n; M; \mathbb{C})$ is a complex Grassmanian, whose points parameterize $n$-dimensional vector subspaces of $\mathbb{C}^{M}$. We will not provide any more details on this approach, referring an interested reader to an excellent textbook~\cite{milnor2016characteristic}. Instead, in this appendix, we will demonstrate the integer nature of $c_{1}$ and $c_{2}$ for the specific and relevant for us cases, considered in section~\ref{sec:conical-topol-inv} by presenting an explicit computation. We also note that in section~\ref{sec:conical-topol-inv} we allowed minor abuse of notation, considering the Chern classes as integer numbers, rather than cohomology classes. The exact proper meaning of Eqs.~(\ref{first-chern-class}) and (\ref{second-chern-class}) is that their l.h.s. represent the Chern classes $c_{1}$ and $c_{2}$, evaluated at the fundamental classes/cycles of $S^{2}$ and $S^{4}$, represented by the identical maps ${\rm id}_{S^{2}}$ and ${\rm id}_{S^{4}}$, respectively.

Explicit expressions for for the closed forms $C_{k} (A)$ that represent the Chern classes are known in a form of a generating function (that generates the classes for all $k$), with the gauge field entering the expressions via its curvature $F(A)$. Here we present the expressions for the first and second classes, relevant for our applications
\begin{eqnarray}
\label{C-one-and-two} &&  C_{1} (A) = \frac{1}{2\pi}{\rm Tr} (F) = \frac{1}{2\pi}{\rm Tr} (F_{ij}) dx^{i} \wedge dx^{j}, \nonumber \\ && C_{2} (A) = \frac{1}{8\pi^{2}}{\rm Tr} (F \wedge F) \nonumber \\ && \;\;\; = \frac{1}{8\pi^{2}}{\rm Tr} (F_{ij} F_{kl}) dx^{i} \wedge dx^{j} \wedge dx^{k} \wedge dx^{l}.
\end{eqnarray}
Note that Eq.~(\ref{C-one-and-two}) represents a local definition, i.e., strictly speaking defines the forms $C_{k, \alpha}$ over $U_{\alpha}$. However, due to the cyclic property of the trace, the connecting/gluing maps for $C_{k, \alpha}$ turn out to be identities, so that we in fact obtain the forms $C_{k}$ defined globally over the whole space $X$.

Verification of the closed nature of $C_{k}$, i.e., checking the conditions $dC_{k} = 0$ for $k = 1, 2$, is a simple and straightforward exercise that involves the properties of the exterior differential operator [Eq.~(\ref{diff-form-prop})], as well as the properties of the trace and commutator. To see independence of $[C_{k} (A)]$ on a particular choice of a gauge field $A$ for the same bundle (gluing data), we note that if $A' = A + a$, then a gauge transformation for $a$ does not have the second (sometimes referred to as inhomogeneous) term in the r.h.s. of Eq.~(\ref{gauge-transform}), i.e., it transforms in the exactly same way as the curvature [Eq.~(\ref{gauge-transform-F})], i.e., $a$ is a globally defined $1$-form with values in the endomorphism bundle. It is another straightforward exercise, which uses the same properties as the previous one, to show
\begin{eqnarray}
\label{C-independ-of-A} C_{1} (A + a) &=& C_{1} (A) + \frac{1}{2\pi} d ({\rm Tr}(a)), \nonumber \\ C_{2} (A + a) &=& C_{2} (A) + \frac{1}{8\pi^{2}} d ({\rm Tr} (a \wedge F)) \nonumber \\ &+& {\cal O} (a^{2}),
\end{eqnarray}
meaning that $C_{k} (A + a)$ differs from $C_{k} (A)$ by an exact form, i.e., the cohomology class $[C_{k} (A)]$ does not depend on a specific choice of a representative, so that the Chern classes $c_{k}$, for $k = 1, 2$, are finally properly defined.

We are now in a position to rationalize Eq.~(\ref{define-degree-S-3}) and derive Eq.~(\ref{second-chern-class}) from Eq.~(\ref{define-degree-S-3}), addressing first the second task. To that end we note that if we denote $U_{\pm} \subset S^{4}$ the contractible subsets of the sphere obtained by withdrawing the north and south poles that correspond to $n_{z} = \pm 1$, respectively, then Eq.~(\ref{sections-connect}) defines a map $g : U_{+} \cap U_{-} \to {\rm SU} (2) \cong {\rm Sp} (1)$ that, being viewed as a gluing data, gives rise to an ${\rm SU} (2)$-bundle over $S^{4}$, with the non-adiabatic terms $A_{\pm}$, defined over $U_{\pm}$, respectively, representing a globally defined gauge field $A$ in the sense explained earlier in this appendix. Therefore the l.h.s. of Eq.~(\ref{second-chern-class}) represents the (integer) value of the second Chern class $c_{2}$ on the fundamental class/cycle of $S^{4}$.

We further proceed with noting that any closed form over any contractible subspace, in particular $C_{2} (A_{\pm})$, is exact. Another straightforward exercise shows that, for $A = A_{\pm}$,
\begin{eqnarray}
\label{C-2-as-exact} && C_{2} (A) = \frac{1}{8\pi^{2}} dB, \nonumber \\ && B = {\rm Tr} (A \wedge dA) + \frac{1}{3} {\rm Tr} (A \wedge [A, \wedge A]) \nonumber \\ && \;\;\; = {\rm Tr} (A \wedge F) - \frac{1}{3} {\rm Tr} (A \wedge A \wedge A),
\end{eqnarray}
and note that $B$ is known in quantum field theory as the Chern-Simons $3$-form. Splitting the integration region $S^{4}$ in Eq.~(\ref{second-chern-class}) into the north and south hemispheres, followed by applying the Stokes theorem to both integrals we obtain
\begin{eqnarray}
\label{C-2-as-degree-g} c_{2} = \frac{1}{8\pi^{2}} \int_{S^{3}} \Delta B, \;\;\; \Delta B = (B_{+} - B_{-})|_{S^{3}},
\end{eqnarray}
with $B_{\pm} = B (A_{\pm})$, and the minus sign in the definition of $\Delta B$ is due to opposite orientations of the hemispheres with respect to the equator $S^{3}$. Also, we again, with a minor abuse of notation, denoted with $c_{2}$ the value $c_{2} ({\rm id}_{S^{4}})$ of the second Chern class on the fundamental cycle of $S^{4}$. Finally upon substitution of
\begin{eqnarray}
\label{C-2-as-degree-g-2} A_{+} = g^{-1} A_{-} g + g^{-1} dg,
\end{eqnarray}
into the second equality in Eq.~(\ref{C-2-as-degree-g}) we obtain after another straightforward computation
\begin{eqnarray}
\label{C-2-as-degree-g-3} && \int_{S^{3}} \Delta{B} = \frac{1}{3} \int_{S^{3}} {\rm Tr} (g^{-1}dg)^{3}, \nonumber \\ &&  (g^{-1}dg)^{3} = g^{-1}dg \wedge g^{-1}dg \wedge g^{-1}dg,
\end{eqnarray}
which completes the derivation.

We conclude this appendix with presenting a more rigorous argument in support of the statement that the degree ${\rm deg} \, g$ of a map $g : S^{3} \to {\rm SU}(2)$ is given by Eq.~(\ref{define-degree-S-3}). It uses a much more invariant definition of the degree of a map $f : S^{n} \to S^{n}$. We first recall that the pullback operation [see Eq.~(\ref{diff-form-pullback})], being applied to closed forms produces a linear map $f^{*} : H^{n} (S^{n}) \to H^{n} (S^{n})$ in the de Rham cohomology. Since, as noted earlier, $H^{n} (S^{n}) = \mathbb{R}$, this linear map is determined by a number, which is called ${\rm deg} \, f$. Since there are integer-valued cohomology theories, e.g., singular or bordism, that stand behind the de Rham real-valued counterpart, the degree is integer valued. We further recognize that the integrand in the r.h.s. of Eq.~(\ref{C-2-as-degree-g-3}) is a pullback $g^{*} \omega$ along $g$ of a $3$-form $\omega$ over ${\rm SU} (2)$, obtained using the same expression by replacing $g$ with the identity map ${\rm id}_{{\rm SU}(2)}$ [for the sake of completeness we note that $\omega$ is a left-invariant form on ${\rm SU} (2)$]. This implies that Eq.~(\ref{C-2-as-degree-g-3}) provides an integral representation for the map degree, if the normalization constant is chosen in such a way so that in case $g = {\rm id}_{{\rm SU}(2)}$ the integral in the r.h.s. of Eq.~(\ref{C-2-as-degree-g-3}) turns to $1$. Therefore, choosing
\begin{eqnarray}
\label{C-2-as-degree-g-4} g (n_{0}, \bm{\sigma}) = \sigma_{0} + i \bm{n} \cdot \bm{\sigma}, \;\;\; n_{0}^{2} + \bm{n}^{2} = 1,
\end{eqnarray}
and performing integration explicitly, e.g., by just using a spherical coordinate system on $S^{3}$, we confirm that Eq.~(\ref{define-degree-S-3}) has the proper normalization constant.

\bibliographystyle{ieeetr}

\end{document}